\documentclass[preprint,12pt]{elsarticle}

\usepackage{hyperref}
\usepackage{array}
\usepackage{graphicx}
\usepackage{siunitx}
\usepackage{multirow}
\usepackage{makecell}
\usepackage{amsmath,amsfonts,amsthm} % Math packages
\usepackage{bm}
\usepackage{subfigure}
\journal{Nuclear Instruments and Methods in Physics Research Section A}

\begin{document}

\begin{frontmatter}

\title{A deep learning approach to multi-track location and orientation in gaseous drift chambers}

%% or include affiliations in footnotes:
\author[mymainaddress]{Pengcheng Ai}

\author[mymainaddress]{Dong Wang}

\author[mymainaddress]{Xiangming Sun\corref{mycorrespondingauthor}}
\ead{sphy2007@126.com}

\author[mymainaddress]{Guangming Huang\corref{mycorrespondingauthor}}
\ead{gmhuang@mail.ccnu.edu.cn}

\cortext[mycorrespondingauthor]{Corresponding authors}

\author[mymainaddress]{Zili Li}

\address[mymainaddress]{Central China Normal University, No.152 Luoyu Road, Wuhan, Hubei, 430079 P.R.China}

\begin{abstract}

Accurate measuring the location and orientation of individual particles in a beam monitoring system is of particular interest to researchers in multiple disciplines. Among feasible methods, gaseous drift chambers with hybrid pixel sensors have the great potential to realize long-term stable measurement with considerable precision. In this paper, we introduce deep learning to analyze patterns in the beam projection image to facilitate three-dimensional reconstruction of particle tracks. We propose an end-to-end neural network based on segmentation and fitting for feature extraction and regression. Two segmentation branches, named binary segmentation and semantic segmentation, perform initial track determination and pixel-track association. Then pixels are assigned to multiple tracks, and a weighted least squares fitting is implemented with full back-propagation. Besides, we introduce a center-angle measure to judge the precision of location and orientation by combining two separate factors. The initial position resolution achieves 8.8 $\mu m$ for the single track and 11.4 $\mu m$ (15.2 $\mu m$) for the 1-3 tracks (1-5 tracks), and the angle resolution achieves 0.15$^{\circ}$ and 0.21$^{\circ}$ (0.29$^{\circ}$) respectively. These results show a significant improvement in accuracy and multi-track compatibility compared to traditional methods.

\end{abstract}


\end{frontmatter}

%\linenumbers

\section{Introduction}
In the context of high energy physics, the \emph{beam physics} studies the characteristics of particles with similar position and momentum in the electromagnetic fields. Charged particle beams (usually generated by synchrotrons) are widely used in many applied and experimental domains. Under many circumstances, measuring the information (position, angle) of individual particles in a beam with high precision has great significance. In large-scale accelerator equipment \cite{Collaboration_2008}, the initial beam undergoes multiple manipulations and its profile becomes very complicated, so it is necessary to accurately measure the particles in the beam. Another example is the hadron therapy \cite{WANG201720} intensely studied in recent years. In order to focus the beam at the location of tumor and form the Bragg Peak \cite{BADANO1999512}, online monitoring the location and orientation of the beam is vital to avoid damaging the healthy body. According to different applications, the required precision of the position varies from millimeter-scale to micrometer-scale.

Based on different detection principles, there are several methods to accurately determine the information of the beam. In \cite{XU2013895} and \cite{Actis_2014}, parallel-plate ionization chambers and strip electrodes were used to measure the ionization charges induced by the incoming particle so as to measure the incident location. Besides, diamond sensors \cite{RN163} using the chemical vapor deposition could tolerate radiation and had high electron and hole mobility, which could be used for beam measurement in extreme environments. In \cite{ALVARADO2020}, array structures made up of the scintillator and photomultiplier tubes (or silicon photomultipliers) could realize measurement with high timing resolution and medium spatial resolution. Finally, in \cite{RN165}, fiber optic radiation sensors transmitting fluorescent light were arranged in a grid to achieve sub-millimeter precision of spatial resolution.

Apart from above methods, silicon-based pixel sensors featuring excellent spatial and timing resolution, rapid response and flexible readout are good candidates in the scenario of beam monitoring. According to the relative location between the pixel sensor and the beam, the detection methods can be divided into two types: \emph{targeting} and \emph{drifting}. The \emph{targeting} type places the pixel sensor facing the incoming beam. The particles in the beam penetrate the pixel sensor and generate charges collected by the electrodes \cite{refId0}. Generally, the \emph{targeting} type has rigorous demands on the radiation hardening, and in most cases the incoming particles are light-weighted, because the deposition of heavy ions will cause the malfunction of the sensor. On the other hand, the \emph{drifting} type utilizes the ionization of beam particles in a gaseous chamber, and an electric field vertical to beam direction forces ionization charges to reach the pixel sensor on the detecting plane. Because the sensor is placed parallel to the beam direction, high energy particles will not penetrate the sensor. Hence, the radiation damage to the sensor is greatly reduced. However, the charge cloud will diffuse in transverse (parallel to the pixel sensor) and longitudinal (vertical to the pixel sensor) directions. How to tackle the diffusion and improve spatial and timing resolution is a main issue for the \emph{drifting} type.

In such a measuring system based on the gaseous drift chamber and the pixel sensor, enabling multi-track location and orientation has a lot of merits (discussed in Section \ref{sec:exp-setup-principle}). In order to obtain the ability, we need to intelligently figure out the presented tracks from an image generated by the pixel sensor for the per-track information. The procedure can be divided into the step of segmentation and the step of regression. For segmentation, much progress has been achieved in recent years with the development of \emph{deep learning} \cite{RN53} techniques. Some representative examples include: the Fully Convolutional Network \cite{7478072} replaced the traditional fully connected layers with convolution layers, and used upsampling to get pixel-level classification results; SegNet \cite{7803544} constructed the encoder-decoder architecture and mapped the index of pooling in the encoder layers to corresponding decoder layers; ENet \cite{DBLP:journals/corr/PaszkeCKC16} worked based on the former two structures, and achieved similar results with less parameters; recently, LaneNet \cite{8500547} adopted the overall architecture from ENet, and used instance segmentation to assist clustering before curve fitting.

In this paper, we present a deep learning architecture to analyze the particular patterns of ionization track projections acquired from the pixel sensor. It is the most important process to reconstruct the three-dimensional (3D) track and get its location and orientation. The main contributions are listed as follows:

\begin{itemize}
	\item We create an end-to-end neural network based on segmentation and fitting. The base network takes the encoder-decoder architecture, and the \emph{binary segmentation} branch and the \emph{semantic segmentation} branch are built upon the base network. The network can effectively extract features from the raw track image, and facilitate subsequent operations.
	\item We invent a \emph{pixel assignment} algorithm to assign the pixels on the weight map to multiple tracks. Afterwards, we implement the \emph{weighted least squares} (WLS) fitting in the software framework of deep learning. The whole network can be optimized end-to-end through back-propagation, and we show the improvement of performance by finetuning the network in an end-to-end way.
	\item To evaluate the results, we propose a \emph{center-angle measure} (CAM) which combines the information from both the location regression and the angle regression. Based on this measure, we investigate the detection rate (for the single track) and F1-scores (for multiple tracks) versus the CAM threshold.
	\item Finally, to demonstrate the practicality of the method, we apply the neural network model to experimental images. The results show the correctness of the simulation process and confirm the same model can generalize to experimental data.
\end{itemize}

\section{Experimental setup and principle}
\label{sec:exp-setup-principle}

Since 2012, the \emph{Topmetal} series \cite{fan2014,AN2016144,Gao_2016} has been developed for beam monitoring applications. The \emph{Topmetal} is a hybrid pixel sensor with position and amplitude resolution. It features ultra low noise level (equivalent noise charge less than 15 electrons) and high charge sensitivity (10-100 \si{\micro\volt}/electron). Around 2016, we applied the test beam using carbon ions with 80.55 MeV/u (mega electron volt per nucleon) energy at Heavy Ion Research Facility in Lanzhou (HIRFL). With the array structure of \emph{Topmetal-II-}, the spatial resolution was better than 20 \si{\micro\meter} and the angle resolution was better than 0.5$^{\circ}$ \cite{WANG201720}. Around 2019, the same pixel sensor was tested using krypton ions with 25 MeV/u at HIRFL. After pre-selection of images, better results were achieved (4 \si{\micro\meter} and 0.15$^{\circ}$) \cite{LI2020163697}. The following sub-sections will first introduce the experimental setup, and then discuss the necessity for multi-track measuring and the proposed multi-track measuring system.

\subsection{Experimental setup}
\label{sec:experimental-setup}

The original beam monitoring system \cite{LI2020163697} was designed to locate single-event latchups of the device under test (DUT). The instrument is the same as Fig. \ref{fig:multi-track-system}, except for the time sensitive region which is not present on \emph{Topmetal-II-}. The locating system is comprised of two independent gaseous drift chambers, each of which is filled with air at 1 atm and room temperature. The DUT is placed on the right side of the second drift chamber. Considering the distance (20 mm) between the two electrodes, the electric field is approximately 250 V/cm in the drift region between the pixel sensor anode (0 V) and the face-to-face metal board cathode (-500 V). In each event, the high-energy ions ionize the gas. Because of the electric field the free electrons then drift towards the anode, where they are collected by the pixel electrodes. The average drift length is 3.2 mm (with a symmetric spread of 1.6 mm) before electrons are collected. The pixel array is a 72 $\times$ 72 matrix with an 83.2 \si{\micro\meter} pixel pitch, which composes a charge-sensitive area of about 6 $\times$ 6 mm$^2$. This is the \emph{Topmetal-II-} specification and will be used as a reference for the new \emph{Topmetal} version in the conceptual design (discussed in Section \ref{sec:multi-system}). Charge signals are read out line by line and pixel by pixel at a working frequency (1.5625 MHz). Finally, track projection images with a fixed refreshing rate (3.3 ms) are sent to and processed by the computer.

Strip-like patterns will appear on the track projection images (dark yellow regions in Fig. \ref{fig:multi-track-system}). The width of the strip is determined by the diffusion coefficients and the drifting distance. The ions experience little multiple scattering. If we assume the electric field is uniform and the ionization charges are distributed ideally, the center line of the strip can represent the direct projection of the traversing path of the ion. One projection line can determine a plane perpendicular to the pixel sensor anode; two such planes have an intersecting line, which can be regarded as the traversing path of the ion.

\subsection{Necessity for multi-track measuring}
\label{sec:necessity}

\begin{figure}[htb]
	\centering
	\includegraphics[width=0.4\textwidth]{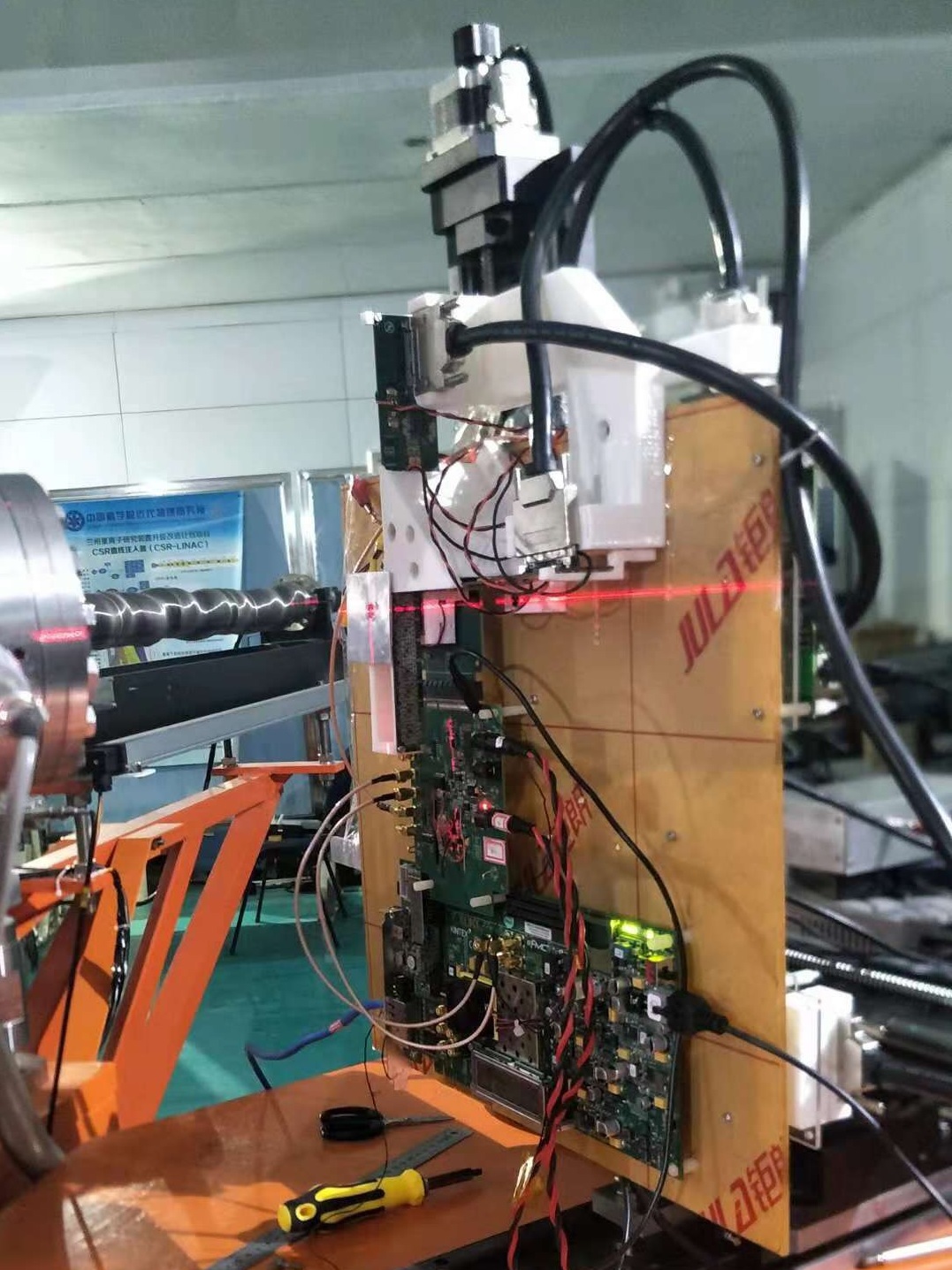}
	\caption{\label{fig:lanzhou} The photograph of the former experiment at HIRFL with the gaseous drift chambers and the \emph{Topmetal} pixel sensor.}
\end{figure}

We applied test beam with krypton ions at HIRFL. A photograph of the testing scene is shown in Fig. \ref{fig:lanzhou}. The flux ranged from 10 ions/(cm$^2\cdot$s) to $1.5 \times 10^4$ ions/(cm$^2\cdot$s). When operating at a low flux, most images from the pixel sensor were empty with background noise, and a few had a single projection track. When we gradually increased the flux, the proportion of single-track images shrank, and the images with two or more tracks became increasingly dominant. According to the above observations, the capability to accurately analyze the multi-track information is vital to improve the efficiency of the detecting instrument. If not, the system must operate at a relatively low flux so that the time needed to reach the required dose of radiation will be inevitably prolonged.

\begin{table}[htb]
	\caption{\label{tab:Poisson} The relation between the quantile of the Poisson distribution and the distribution parameter $\lambda$ at different values of detection efficiency. The flux (ions/(cm$^2\cdot$s)) is estimated according to the experimental setup. Probability distribution function: $f(k)=exp(-\lambda)\lambda^k/(k!)$}
	\centering
	\scriptsize
	%% \tablesize{} %% You can specify the fontsize here, e.g., \tablesize{\footnotesize}. If commented out \small will be used.
	\begin{tabular}{c|cccccc}
		\hline
		\multirow{2}{*}{\textbf{Quantile}}	& \multicolumn{2}{c}{\textbf{95\% detection}}	& \multicolumn{2}{c}{\textbf{97\% detection}} & \multicolumn{2}{c}{\textbf{99\% detection}} \\
		\cline{2-7}
		& $\lambda$ & flux & $\lambda$ & flux & $\lambda$ & flux \\
		\hline
		k=1	& 0.3554 & 299.2 & 0.2676 & 225.3 & 0.1486 & 125.1 \\
		k=2 & 0.8177 & 688.3 & 0.6649 & 559.7 & 0.4361 & 367.1 \\
		k=3 & 1.3664 & 1150 & 1.1551 & 972.3 & 0.8233 & 693.0 \\
		k=4 & 1.9702 & 1658 & 1.7061 & 1436 & 1.2792 & 1077 \\
		k=5 & 2.6131 & 2200 & 2.3005 & 1936 & 1.7853 & 1503 \\
		\hline
	\end{tabular}
\end{table}

Assuming the incoming ions in a unit of time obey the Poisson distribution, we could argue that the distribution parameter $\lambda$ is proportional to the flux. Table \ref{tab:Poisson} shows the relation between the quantile and distribution parameter $\lambda$ at different values of detection efficiency. It can be seen that when the detection efficiency is fixed, $\lambda$ will increase with respect to the increase of the quantile, and the tendency gradually becomes speeding up. This phenomenon is more obvious at high detection efficiency. We might take the first line and the third line in the table as an example. When the values of detection efficiency are 95\%, 97\%, 99\%, the $\lambda$ values when k=3 are 3.84, 4.32, 5.54 times bigger than those when k=1. It demonstrates that increasing recognizable multiple tracks will make it possible for the system to work at higher fluxes. Meanwhile, the decrease of empty background images could also improve the efficiency of the data transmission back-end.

\subsection{Multi-track measuring system}
\label{sec:multi-system}

\begin{figure}[htb]
	\centering
	\includegraphics[width=0.95\textwidth]{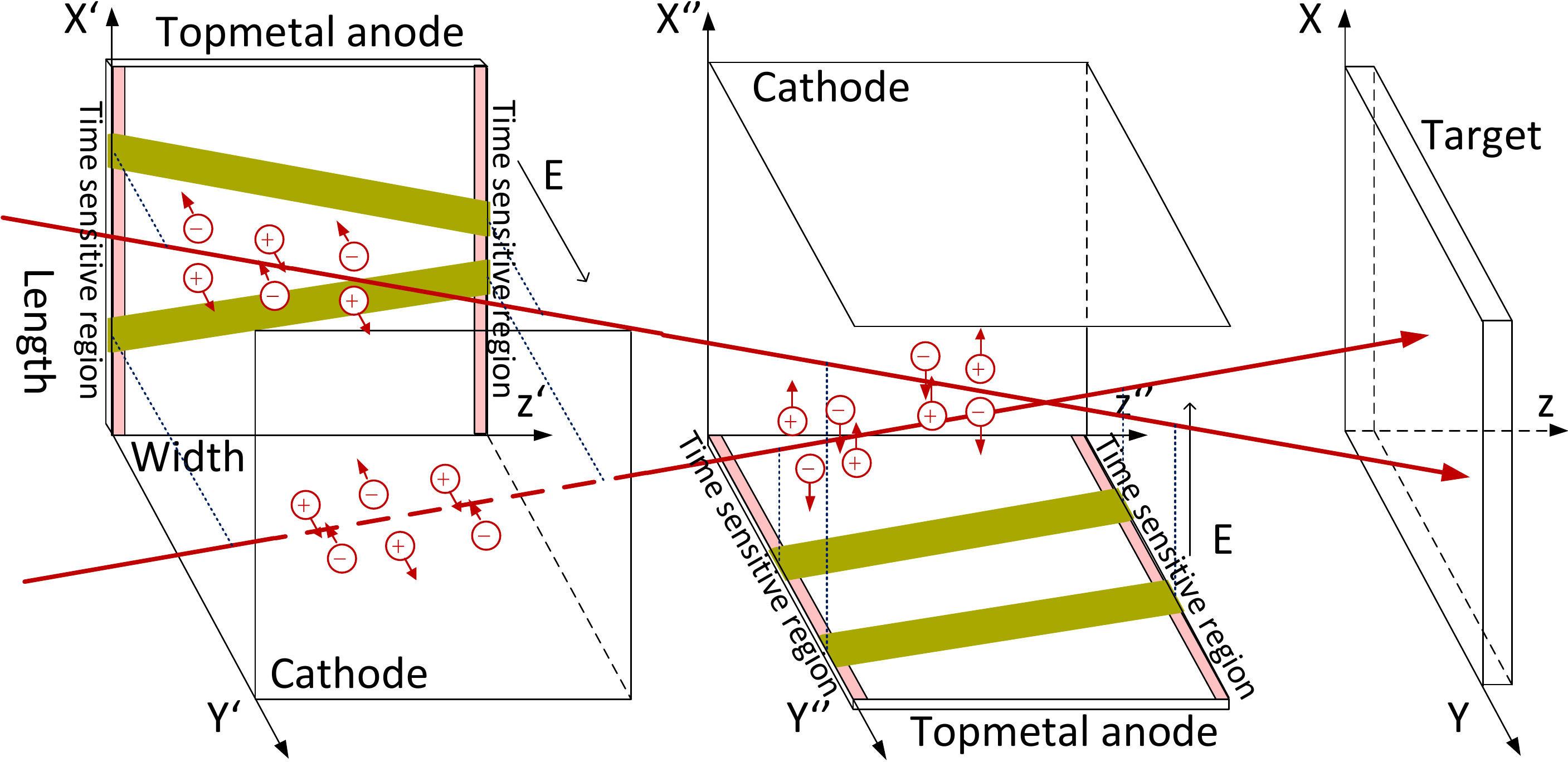}
	\caption{\label{fig:multi-track-system} A conceptual design of the multi-track measurement system enabled by the \emph{Topmetal-CEE} pixel sensor. This figure shows two incoming particles and their ionization track projections onto the \emph{Topmetal-CEE} chip. The \emph{Topmetal-CEE} provides position \& amplitude information (the square area) and time information (the time sensitive region).}
\end{figure}

To implement the detecting scheme, the issue of ambiguity needs to be settled when reconstructing the 3D track from coordinate plane projections. Since each track projection could determine a perpendicular plane, if there are several tracks projecting on the two coordinate planes (assuming n tracks), the reconstructed 3D tracks have more than one possibility (total n!) following the law of combination. To eliminate the ambiguity, there are some feasible methods, such as adding a third projection plane, setting a reference point outside the detecting instrument, building extra detectors for coincidence measurement, etc. Beyond these methods, we would like to solve the problem with the pixel sensor alone. This is why we propose to design \emph{Topmetal-CEE} specially optimized for the Cooler Storage Ring External Target Facility Experiment (CEE) \cite{RN169} in the near future. Fig. \ref{fig:multi-track-system} shows the conceptual design of the detecting system enabled by \emph{Topmetal-CEE}. This system could be applied to beam monitoring in various circumstances when a target is in the reasonable range. The major innovation compared to \cite{LI2020163697} is the \emph{time sensitive region} integrated in the chip. This region has sufficient timing ability (better than 1 \si{\micro\second}) and reasonable position resolution (depending on the pixel size, usually $\sim$100 \si{\micro\meter}). When a high energy ion passes through the drift chamber, the drifting charges will trigger the time sensitive region on both sides of the pixel sensor. The timing information is thus recorded and buffered inside the chip. At fixed reset points, the recorded times are read out and used to match the tracks projecting on different coordinate planes.

\begin{figure}[htbp]
	\centering
	\includegraphics[width=0.95\textwidth]{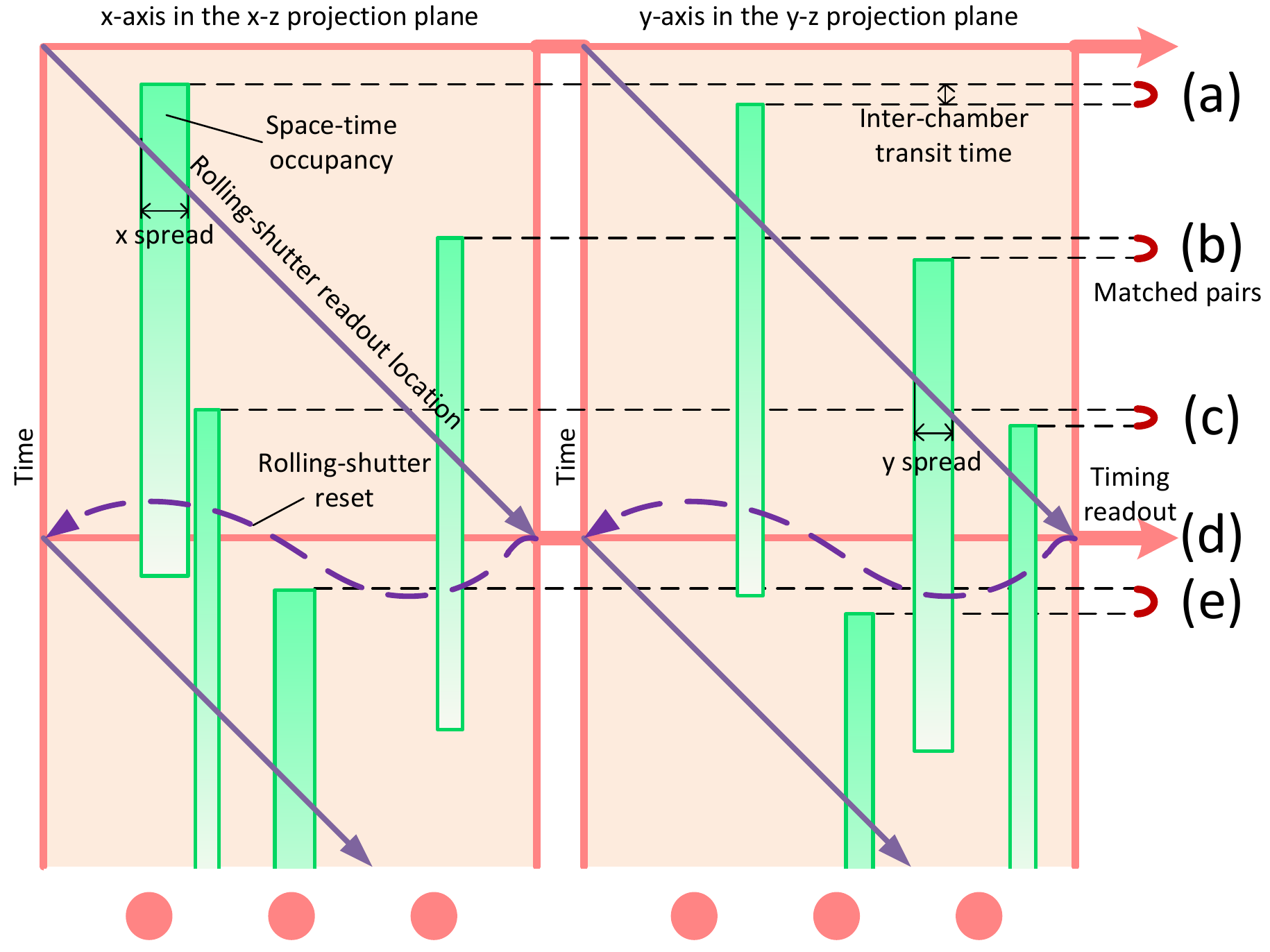}
	\caption{\label{fig:timeline} The timeline of the readout scheme. (a) The first ion enters the detection area and leaves tracks on the x-z plane and the y-z plane, and the crossing times are recorded by time sensitive regions of \emph{Topmetal-CEE} so as to match the tracks. (b) The second ion enters the detection area and the tracks overlie the previous; however it is still distinguishable because times recorded by two planes can be matched. (c) The third ion enters, and it is beyond the subsequent shutters of the x-z plane while remaining in the subsequent shutters of the y-z plane; besides, times are recorded. (d) At the reset point, recorded times are read out, and the shutters are reset. (e) The next ion enters, and the track overlies the previous on the x-z plane.}
\end{figure}

Fig. \ref{fig:timeline} shows a typical timeline of the readout scheme. The main complexities arise from the row-major order (iterating in a row first, then row-by-row) of the rolling-shutter readout. We assume synchronization is maintained between the two pixel sensors, and the spatial and amplitude readout is reset at periodical reset points (only the read pointer is reset, not the signals). Since the locations of the incoming ions are completely random, it is possible to receive the track projection in the current interval but read out in the next interval (x-z plane in Fig. \ref{fig:timeline}). As a result, the times read out at a reset point might correspond to a track projection in the following interval; besides, the same ion might project on the two coordinate planes in two consecutive intervals. In spite of these complexities, in most cases the amplitude readout can match the recorded times, and times between two coordinate plane are well associated according to the kinetics of the incoming ion. We only need to enlarge the time-space matching region in a single pixel sensor from the current interval to consecutive two intervals. As long as the multi-track location and orientation is accurate in a single image, the 3D reconstruction is achievable with timing information. Besides, information-rich projection images facilitate high-precision measurement.

A possible issue is the partial track images due to the rolling-shutter readout. According to the geometric structure, the direction of the detected ion deviates from the z-axis little, so partial tracks are not significant considering the row-major readout. Besides, if a partial track does happen in the current interval, by carefully designing the discharge circuit in each pixel, we can control the discharge time to be approximately (and optimally) one period. This makes the system always capture at least one whole track in two successive intervals without counting a track several times. Although switching to global-shutter readout or using multiple parallel output buffers might be more robust solutions, we believe the proposed scheme can fix the issue of partial tracks in most cases.

\section{Empirical methods and their limitations}
\label{sec:empirical}

In two-dimensional images within the subsequent sections, if not specified, the horizontal axis and the vertical axis represent indexes of the pixels in \emph{Topmetal-II-}. The same method is applicable to other reasonable sizes in the proper experimental condition.

\subsection{Mass center method}

\begin{figure}[htb]
	\centering
	\includegraphics[width=0.75\textwidth]{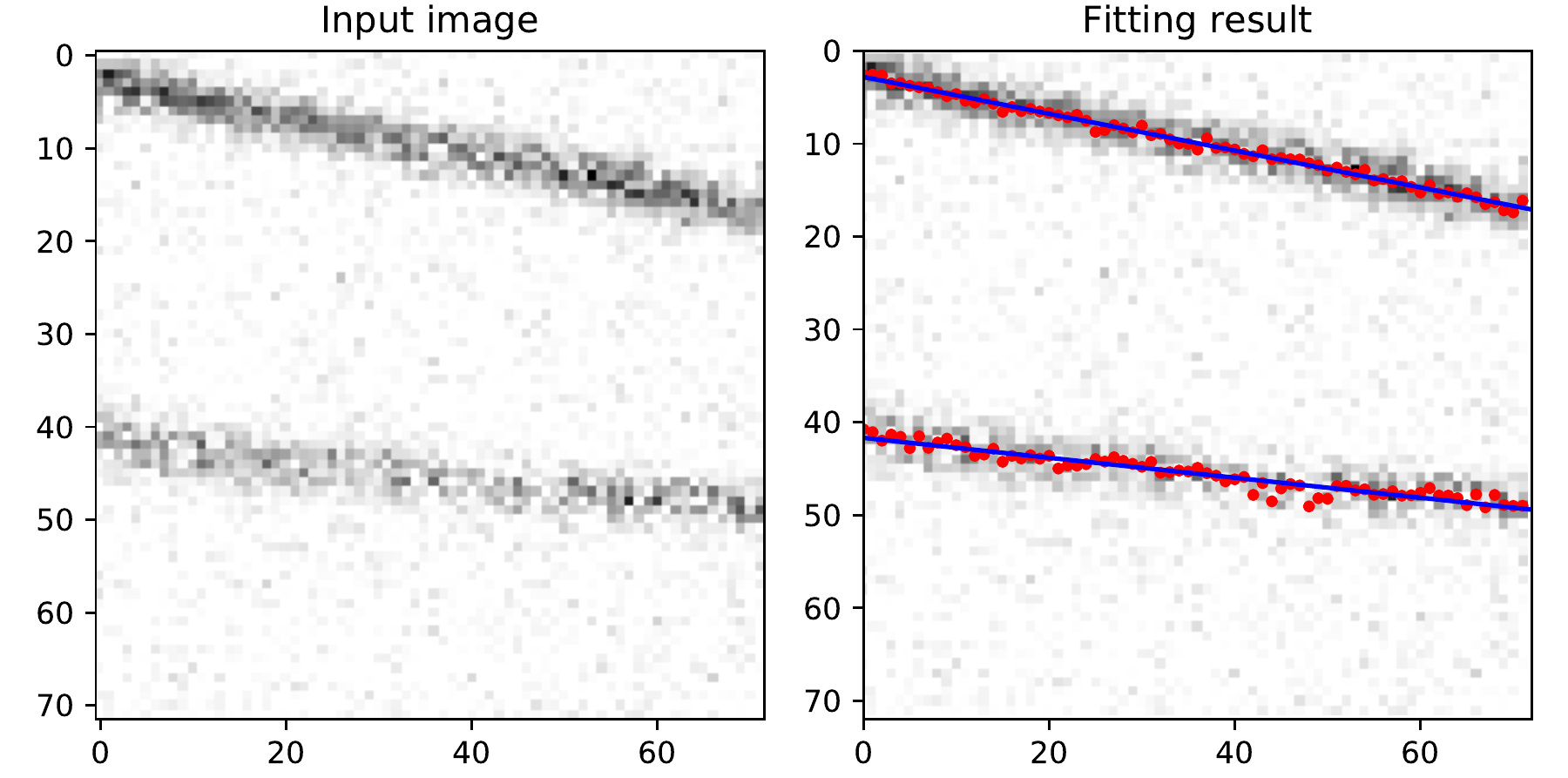}
	\caption{\label{fig:mass-center-example} Location and orientation with the mass center method. (left) There are two tracks on the input image. (right) We manually separate the upper part of the image and the lower part of the image, and use the mass center method individually.}
\end{figure}

In the ideal condition (straight line and symmetric distribution), the center line of the track images could represent the projection of the traversing path. \emph{Topmetal} is a device collecting ionization charges directly without gas-electron multiplication. Each pixel electrode collects drifting electrons and charges the capacitor. To utilize the amplitude readout, in \cite{LI2020163697} the \emph{mass center method} is used to fit the position and direction of the tracks, which is illustrated in Fig. \ref{fig:mass-center-example}. When there is only one track in the fitting region, the fitting procedure is straightforward. We find the pixel with maximum amplitude in a column as the reference point. Then upper and lower pixels are selected along with the center pixel (5 in either side, 11 in total), and a mass center is calculated with the amplitude as the mass of each chosen pixel. This procedure is repeated for each column. Finally, a straight line is fitted according to the position of the mass centers.

The mass center method is based on the assumption of the symmetric charge distribution, and takes advantage of amplitude readout of the hybrid pixel sensor. It is used in practice successfully. When the diffusion of the electron cloud is not so significant, the efficiency of the charge collection is relatively high and the integrity of projection tracks is guaranteed, the mass center method has good accuracy (see Section \ref{sec:single-sim}).

\subsection{Double edge detection method}

\begin{figure}[htb]
	\centering
	\includegraphics[width=0.75\textwidth]{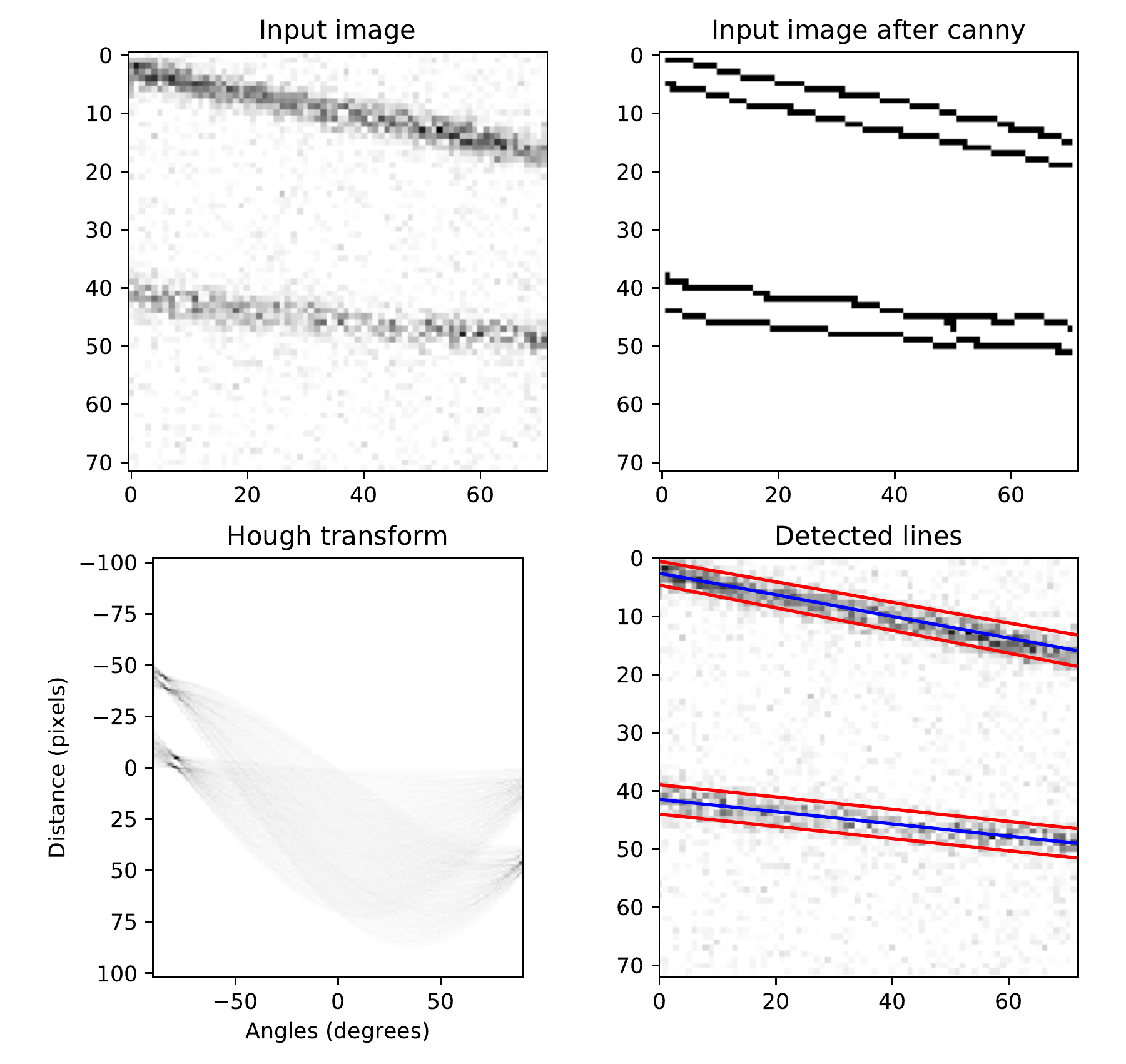}
	\caption{\label{fig:double-edge-example} Location and orientation with the double edge detection method. (left top) The input image is the same as Fig. \ref{fig:mass-center-example}. (right top) Canny edge detection algorithm is applied to the input image. (left bottom) Hough transform converts the edge image to the Hough space. (right bottom) Straight lines in the Hough space are matched, and the center line of each matched pair is regarded as the projection of the traversing path.}
\end{figure}

Different from the mass center method which utilizes the intensity of the input image, the \emph{double edge detection method} works with the contrast at the edge locations. Since the electron cloud forms patterns with a certain width on the image sensor, there are obvious changes of the intensity at the two edges of the strip. The procedure is shown in Fig. \ref{fig:double-edge-example}. First, we process the input image with the Canny edge detection algorithm \cite{4767851}. Then a Hough transform is performed with the edge image so that distances and angles of the edge lines are detected and matched. Finally, the center line of each matched pair is regarded as the projection of the traversing path.

The double edge detection method mainly uses the features of edges on the track projection image. Its computations are based on the image after the edge detection algorithm. When the structures of the tracks are integral and multiple tracks are far away at a certain distance, the double edge detection method could detect multiple tracks and give the information of each track separately. Its performance on single-track images is evaluated in Section \ref{sec:single-sim}.

\subsection{Limitations of empirical methods}

The above two empirical methods could achieve satisfactory results in proper conditions. However, some issues with these methods are hard to resolve in practice. Here we make a discussion about the limitations of each method.

\paragraph{Mass center method} The original algorithm of the mass center method targets the single track case. When there are multiple tracks on the projection image, it does not have the ability to infer the number of tracks. An inaccurate number will cause large deviations of the results, especially when there are 3 or more tracks. Furthermore, even if we can design an effective rule to judge the number of tracks before mass center calculation, the amplitude of two tracks will interplay at the overlapping regions, and the final precision will deteriorate.

\paragraph{Double edge detection method} The practical issue of the double edge detection method is how to determine the parameters in the algorithm. The Hough transform requires the divisions of the distance and the angle. If the number of divisions is too small, the "pixel" in the Hough space will be too large and it will lower the precision; if the number of divisions is too large, the accumulations in each "pixel" of the Hough space will decrease, which will hamper non-maximum suppression (NMS). Besides, if parameters are not chosen properly in the process of NMS, either some edges will be missed, or pseudo edges will present in the result. Finally, it is hard to find universal parameters that work well with a dataset of images, or images continuously acquired from the detecting instrument.

As a supplement, these two methods belong to empirical methods based on some prior assumptions of the ideal conditions. However, the physical process is complex and conditions out of those assumptions will happen. For example, if there is a "bright point" in the projection image due to the abnormal status of the pixel sensor, the mass center method will mistakenly treat it as the reference point. In conclusion, empirical methods have their safe regions and behave inadequately to account for real-world complexities.

\section{Architecture}
\label{sec:arch}

According to the analysis in Section \ref{sec:empirical}, empirical methods use fixed routines to infer the position and the direction of tracks. There is still much room to improve by utilizing the information in the image thoroughly and handling multiple tracks in a single image reliably. In particular, deep learning techniques based on convolutional neural networks (CNN) could extract features from the input data at different levels. At the front layers, shallow features (intensity, line segment, angle) could be extracted; when the network goes deep, the pixels in the feature map have increasingly large receptive fields, and the features are more abstract (pattern or instance). Due to the hierarchy of the neural networks, they can find out the underlying relations between the input data and finish advanced classification or regression tasks.

The definition of beams in high energy physics indicates that most particles in the beam have similar location and momentum. To make our neural network model work in more general scenarios, we actually consider the multi-track problem; the initial positions and directions of the particles change in a certain range. This actually raises the difficulty to obtain the information of tracks individually. In the experiments (Section \ref{sec:exp-images}), it can be seen that the same neural network model could work very well in the real-world track projection images.

In the multi-track problem, we need to judge the number of tracks from the input projection image, and measure each track accurately. It is not only a one-to-many problem, but also an indeterminate problem. Furthermore, the principle of the beam line implies that each particle in the beam is independently and identically distributed, which means no prior knowledge about the relative locations of multiple tracks. However, traditional CNN structures construct a fixed mapping between the input data and the output class/variable, and the target of CNN is a label with predefined dimensions. This makes them inflexible to accommodate varied targets in the problem. Hence, we need to explore new methods to discover independent modes in the projection images and possess the regression ability at the same time.

\begin{figure}[htbp]
	\centering
	\includegraphics[width=0.95\textwidth]{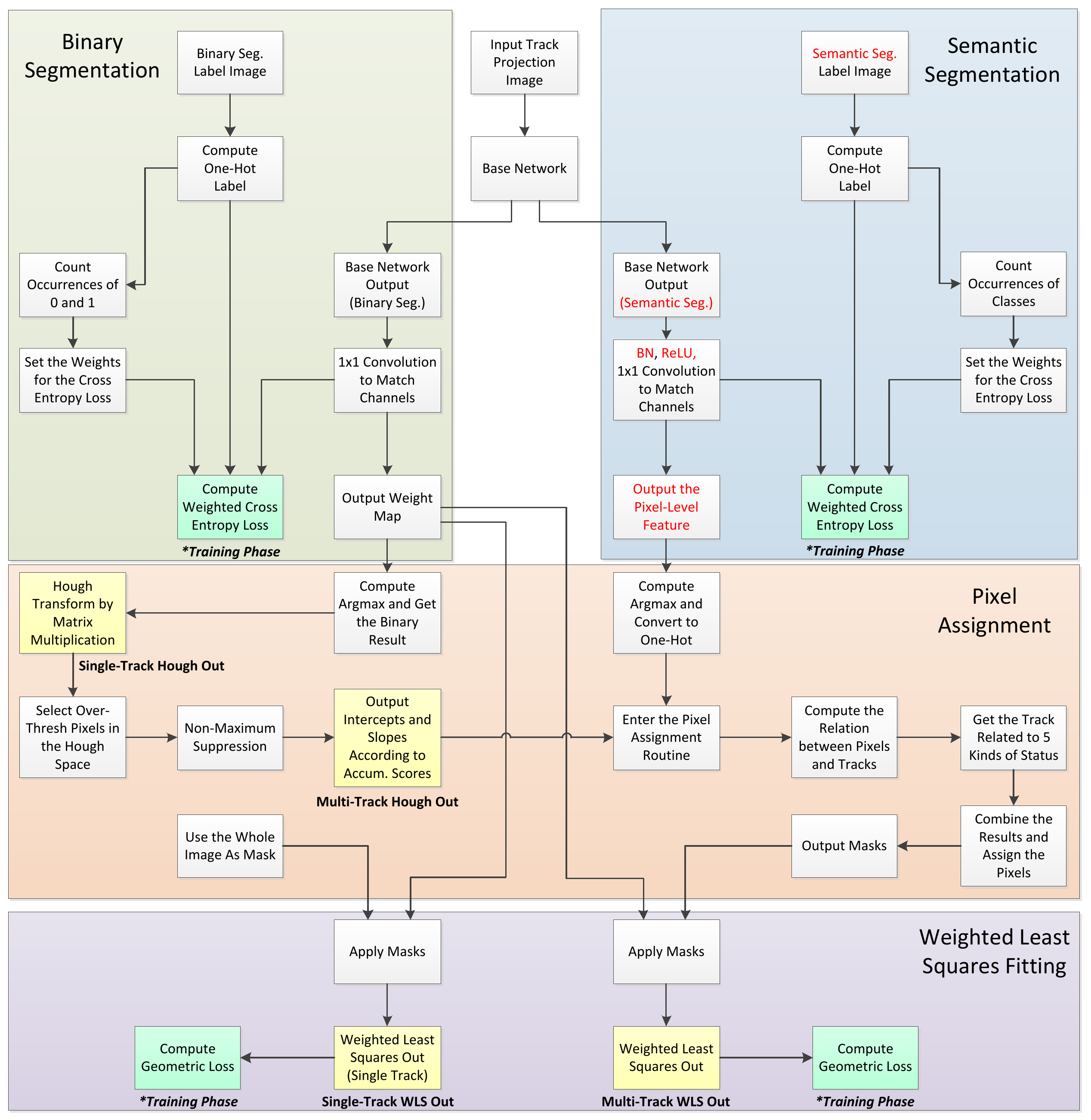}
	\caption{\label{fig:seg-graph-combine} The overall architecture of the proposed network model. The base network is introduced in Section \ref{sec:base-network}. Four shaded sub-diagrams are described in subsequent sub-sections. For the single-track case, pixel assignment is not necessary for the WLS fitting. For the multi-track case, pixel assignment combines the two branches and provides the mask for each track. The green squares represent loss functions used for training. The yellow squares represent outputs of the network analyzed in the simulation and experiment.}
\end{figure}

To tackle the multi-track problem, we create an end-to-end neural network based on segmentation and fitting. The overall architecture of the network is shown in Fig. \ref{fig:seg-graph-combine}, which will be explained later in detail. The base network, the binary segmentation, the semantic segmentation with pixel assignment, and the WLS fitting will be described separately. Finally, the configurations of the overall architecture are introduced.

\subsection{Base network}
\label{sec:base-network}

\begin{figure}[htbp]
	\centering
	\includegraphics[width=0.95\textwidth]{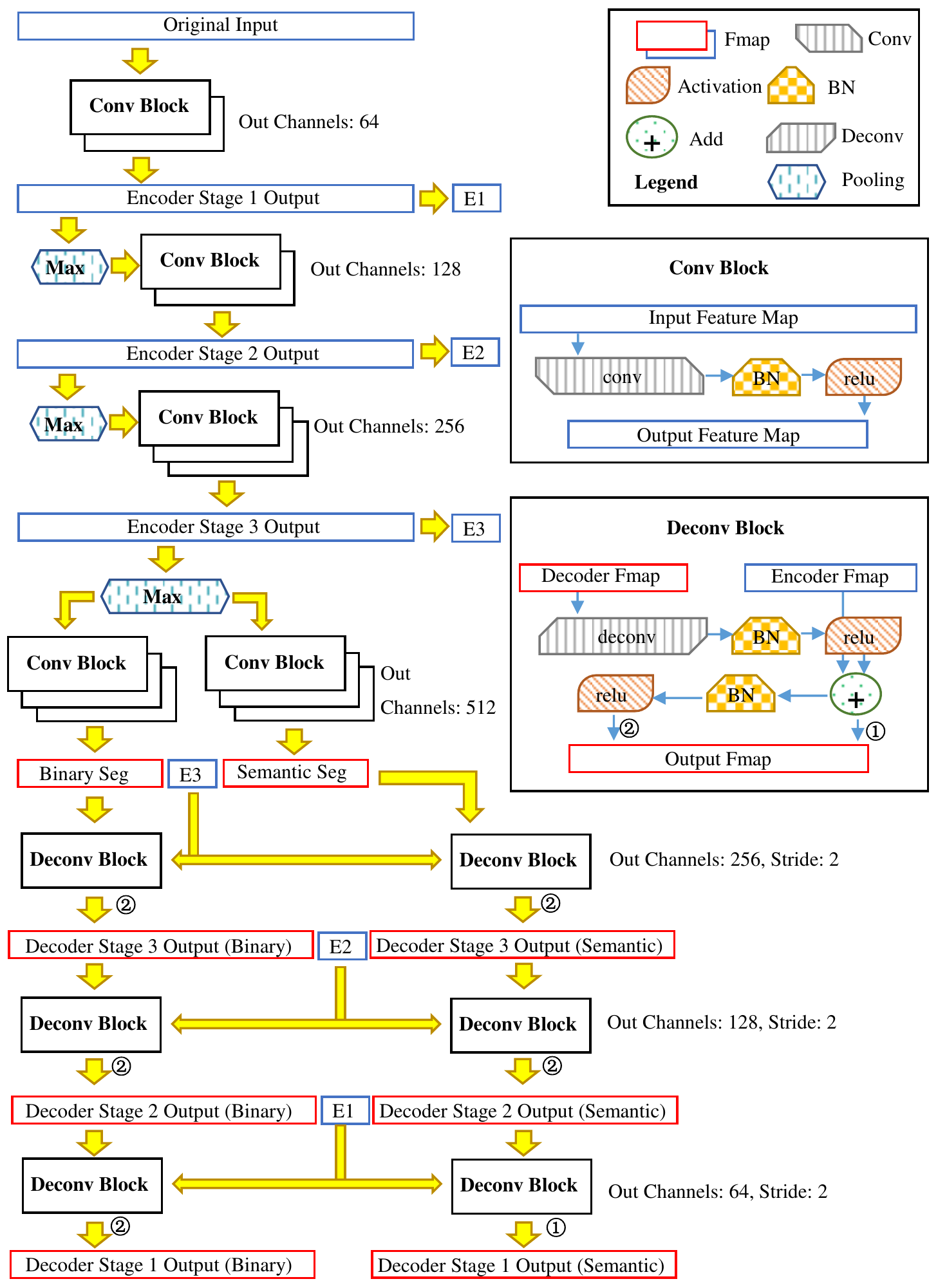}
	\caption{\label{fig:base-network-arch} Base network of the proposed network architecture.}
\end{figure}

The base network in the architecture is shown in Fig. \ref{fig:base-network-arch}. The configuration of parameters in each layer refers to the VGG Net \cite{RN81}. The base network is made up of the encoder part and the decoder part. Convolutions with kernel size 3 $\times$ 3 and stride 1 are used in the encoder. The \emph{same} padding along with the unit stride implies that the size of feature maps will not change in convolution layers. The convolution, the batch normalization \cite{10.5555/3045118.3045167} and the rectified linear unit (ReLU) activation form the Convolution Block (Conv Block). The encoder contains three stages. In the first stage, there are two successive convolution blocks, and the number of output channels is 64. In the second stage and the third stage, the feature maps will first go through a max pooling layer with window size 2 $\times$ 2 and stride 2. Successive convolution blocks follow with 128 output channels in the second stage and 256 output channels in the third stage. The output of the third stage will go through a similar max pooling layer and then divide into two branches. Each branch has three successive convolution blocks. We name them the binary segmentation branch and the semantic segmentation branch in the subsequent decoder part.

In the decoder part, deconvolutions are used. If we denote the convolution operation with the matrix multiplication of the input feature map and a sparse matrix, the deconvolution can be viewed as the matrix multiplication of the output feature map and the transpose of the sparse matrix. The deconvolutions have kernel size 4 $\times$ 4, stride 2 and \emph{same} padding, so the feature map is twice larger after them. The deconvolution and relevant operations form the Deconvolution Block (Deconv Block). In the deconvolution block, the feature map in the decoder will undergo deconvolution, batch normalization and ReLU; after those operations, the result is added by the corresponding feature map in the encoder, which resembles the structure in ResNet \cite{7780459}. The added result can be sent out directly or after another batch normalization and ReLU. The decoder has three stages with reverse sequence numbering. The binary segmentation branch and the semantic segmentation branch share the encoder and have their own decoder layers. The output channels in the decoder correspond to the encoder (256, 128, 64). All deconvolution blocks output from \textcircled{2} except the last stage in the semantic segmentation branch.

\subsection{Binary segmentation}

In deep learning and computer vision, the binary segmentation usually refers to generating a feature map with the same size as the original input; each pixel on the feature map could be 0 or 1 so as to discriminate a single class or a single instance. In our multi-track problem, we generalize the concept to also include the weight map before argmax (argument of a function when maximum) operation to get the binary map.

For the single-track case, the function of the binary segmentation is to determine the center line of the track to facilitate Hough transform; besides, we can use the weight map before binarization to make a WLS fitting. For the multi-track case, the binary result can also be used for coarse location and orientation through Hough transform; or we can combine the weight map with the semantic segmentation and assign each pixel to different tracks. After that, the WLS fitting can be performed to provide accurate information of multiple tracks.

The flow diagram of the binary segmentation process is shown in light green shade in Fig. \ref{fig:seg-graph-combine}. The binary segmentation output in the base network has 64 channels. To fit into the two binary classes, a convolution layer with kernel size 1 $\times$ 1 and stride 1 is used to match the number of channels. In the test phase, argmax can be computed upon the convolution.

In the training phase, a label image for the binary segmentation is sent into the network and converted to the one-hot format (a vector with "1" in a single entry and "0" elsewhere) along the depth dimension. Since the label classes are extremely unbalanced (only the center line of each track is marked with "1"), original cross entropy loss will be biased towards backgrounds. Hence it is necessary to set larger weights for the "1" class to cancel the bias. To achieve this, we count the occurrences of "0" and "1" in a minibatch and set the weights inversely correlated with the frequency:

\begin{equation}
\label{equ:weight-for-ce}
w_i = \frac{1}{ \log(\frac{N_i}{\sum_j N_j} + \epsilon) }
\end{equation}

\noindent where $N_i$ is the number of occurrences of class $i$, and $\epsilon$ is a positive number slightly bigger than 1 (we set it to 1.02), and $w_i$ is the weight for class $i$.

To calculate the weighted cross entropy, softmax is applied to the dimension of channel in the feature map after 1 $\times$ 1 convolution. Then the following loss is computed:

\begin{equation}
\label{equ:softmax}
\text{softmax}(\bm{x})_i = \frac{\exp(x_i)}{\sum_{j=1}^n \exp(x_j)}
\end{equation}

\begin{equation}
\label{equ:weighted-ce}
L = -\sum\limits_{m,n} \sum\limits_i \bm{1}(y_{m,n}=i) \cdot w_i \log(\text{softmax}( \bm{x}_{m,n} )_i )
\end{equation}

\noindent where $\bm{x}_{m,n}$ is the vector at (m, n) in the feature map, softmax is computed along the vector. $\bm{1}(\cdot)$ is the indicative function which takes 1 when the condition is satisfied, and 0 otherwise. $y_{m,n}$ is the label at (m, n) indicating the ground-truth class. The binary segmentation branch can be trained with the weighted cross entropy loss by back-propagation.

\subsection{Semantic segmentation with pixel assignment}

In computer vision, the semantic segmentation usually refers to classifying the pixels in an image according to its class property. In our multi-track problem, the task for the semantic segmentation and the following pixel assignment is to assign the pixels of the weight map to the most appropriate tracks so as to use the WLS fitting for accurate location and orientation.

\begin{figure}[htb]
	\centering
	\includegraphics[width=0.85\textwidth]{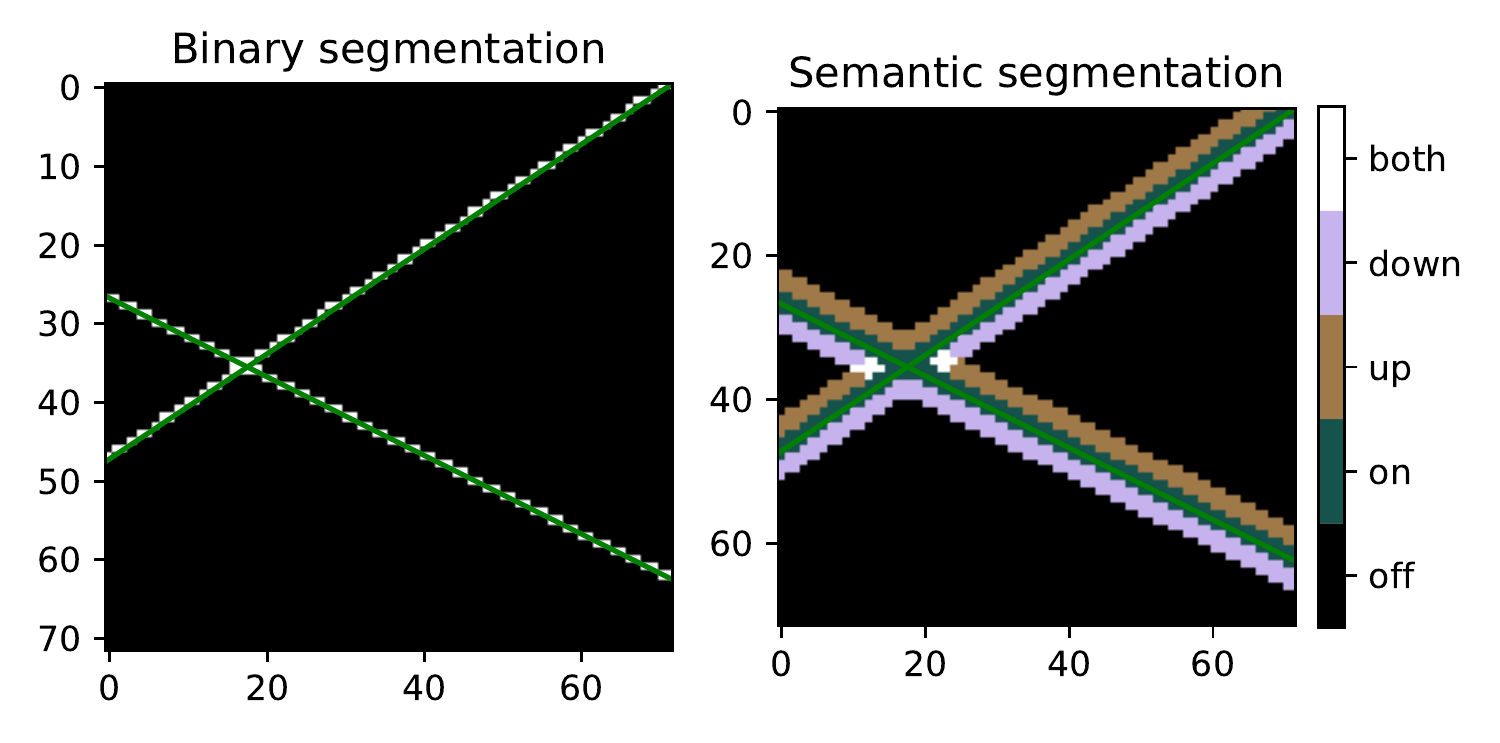}
	\caption{\label{fig:seg-example} An example to show how binary segmentation and semantic segmentation work. We set the angles of tracks with regard to the horizontal axis more obvious than reality for better visualization. (left) Binary segmentation gives the center lines of tracks as well as a weight map for weights in the WLS fitting. (right) Semantic segmentation annotates each pixel to one of five relative locations, and the pixel assignment uses the annotations and the Hough transform of the binary map to assign pixels to proper tracks. After that, the WLS fitting can be performed.}
\end{figure}

Unlike what is normally considered, the semantic segmentation in our multi-track problem is to divide the pixels according to relative locations. To be more specific, we classify each pixel to one of five kinds of status: \emph{Off}, \emph{On}, \emph{Up}, \emph{Down} and \emph{Both}. The method in detail is shown in Fig. \ref{fig:seg-example}. In the training phase of semantic segmentation, we set label images related to track projection images to optimize the network parameters. In the test phase, the status of each pixel is predicted. Before the WLS fitting, we use operators inside the deep learning framework to assign pixels on the weight map to proper tracks according to the results of semantic segmentation.

The flow diagram of the semantic segmentation process is shown in light blue shade in Fig. \ref{fig:seg-graph-combine}. The diagram is similar to binary segmentation, except for the following facts: (1) batch normalization and ReLU are used before 1 $\times$ 1 convolution; (2) Output the pixel-level feature instead of the weight map before pixel assignment; and (3) the number of matched channels or the classes increase from 2 to 5. When computing the weighted cross entropy loss, we still use Equation (\ref{equ:weight-for-ce}), Equation (\ref{equ:softmax}) and Equation (\ref{equ:weighted-ce}). The only difference is the number of channels.

The pixel assignment process combines the binary segmentation and the semantic segmentation for the purpose of WLS fitting, which is shown in light orange shade in Fig. \ref{fig:seg-graph-combine}. The first step is to map the results of binary segmentation to the Hough space to roughly judge the number and location of tracks. In order to implement the Hough transform inside the deep learning framework, the following element-wise matrix multiplication is used:

\begin{equation}
H(d, \theta) = \sum\limits_{m,n} C(m,n,d,\theta) \cdot P(m,n)
\end{equation}

\noindent where $C$ represents the conversion matrix of the Hough transform, $P$ is the binary map and $H$ is the result of the Hough transform.

Next, we select the significant regions in the Hough space by setting a threshold to half the maximum accumulations, and invoke the NMS afterwards. The outputs of NMS are converted into (intercept, slope) pairs, sorted according to the intercepts, and sent to the pixel assignment routine. On the other hand, argmax is applied to the pixel-level feature of semantic segmentation, and the one-hot results are also transferred to the pixel assignment routine.

In the pixel assignment routine, we compute the following arguments of extrema:

\begin{equation}
d(i,m,n) = k_i \cdot m - n + b_i
\end{equation}

\begin{gather}
\text{nn}(m,n) = \mathop{\arg\min}\limits_{i} d(i,m,n) \quad s.t. \quad d(i,m,n) \geq 0 \\
\text{pos}(m,n)= \mathop{\arg\min}\limits_{i} d(i,m,n) \quad s.t. \quad d(i,m,n) > 0 \\
\text{np}(m,n) = \mathop{\arg\max}\limits_{i} d(i,m,n) \quad s.t. \quad d(i,m,n) \leq 0 \\
\text{abs}(m,n)= \mathop{\arg\min}\limits_{i} |d(i,m,n)|
\end{gather}

\noindent where $k_i$ represents slope, $b_i$ represents intercept, and $d$ represents the vertical distance between point (m,n) and the i-th track center line. nn, pos, np and abs represent the track indexes of the nearest non-negative, the nearest positive, the nearest non-positive and the nearest absolute with regard to point (m,n). If no track satisfies the constraints, the value will be -1 (invalid). Then, the masks are generated as follows:

\begin{equation}
\text{mask}(i,m,n) = \left\{
\begin{array}{rcl}
1, & & if \ \text{abs}(m,n) = i \ and \ I(m,n,1) = 1 \\
1, & & if \ \text{nn}(m,n) = i \ and \ I(m,n,2) = 1 \\
1, & & if \ \text{np}(m,n) = i \ and \ I(m,n,3) = 1 \\
1, & & if \ (\text{pos}(m,n) = i \ or \ \text{np}(m,n) = i) \ and \ I(m,n,4) = 1 \\
0, & & otherwise
\end{array}
\right.
\end{equation}

\noindent where $I$ represents the one-hot predictions of semantic segmentation, and the mask represents the assigned pixels corresponding to each track.

\begin{figure}[htb]
	\centering
	\includegraphics[width=0.8\textwidth]{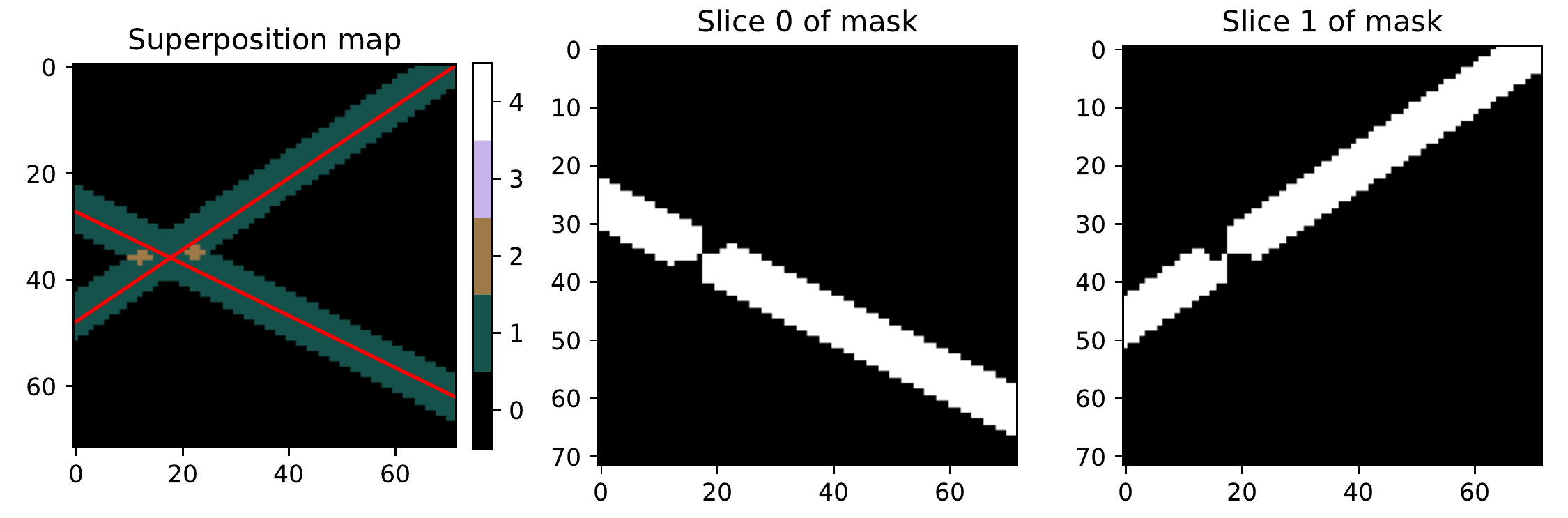}
	\caption{\label{fig:inst-seg-with-tf} The pixel assignment based on the binary segmentation and semantic segmentation in Fig. \ref{fig:seg-example}. (left) Superposition map of all the assigned pixels. (center) The pixels assigned to the first track. (right) The pixels assigned to the second track.}
\end{figure}

The example of the pixel assignment is shown in Fig. \ref{fig:inst-seg-with-tf}. It can be seen that for most pixels the two tracks are discriminated very well. At the intersection of the two tracks, there will be a "gap" because of the mutual effect in the pixel assignment. However, it is not significant when we take the overall integrity into consideration. In the training phase, the WLS fitting is performed and the residuals are back-propagated. The optimization process ensures that the representative features are learned and the negative effect of the "gap" is reduced.

\subsection{Weighted least squares fitting and its back-propagation}

With the pixel assignment results, we can use the weight map before binarization to make a weighted least squares (WLS) fitting, which is shown in light purple shade in Fig. \ref{fig:seg-graph-combine}. Consider the following system of linear equations:

\begin{gather}
\bm{X}\bm{\beta} = \bm{Y} \label{equ:linear-system} \\
where \quad \bm{X} =
\begin{pmatrix}
1 & x_1 \\
1 & x_2 \\
\vdots & \vdots \\
1 & x_m 
\end{pmatrix}, \quad
\bm{\beta} =
\begin{pmatrix}
\beta_1 \\
\beta_2
\end{pmatrix}, \quad
\bm{Y} =
\begin{pmatrix}
y_1 \\
y_2 \\
\vdots \\
y_m
\end{pmatrix} \nonumber
\end{gather}

\noindent where $x_i$ and $y_i$ are the abscissa and the ordinate of the i-th pixel involved in the fitting, and $\beta_i$ is the i-th parameter to be fitted. Usually $m \gg 2$, so the above linear equations are over-determined and no unique solution exists in the common case. To achieve the best solution, the following optimization problem needs to be solved:

\begin{equation}
\bm{\beta} = \mathop{\arg\min}\limits_{\bm{Z} \in R^{2 \times 1}} || \bm{X} \bm{Z} - \bm{Y} ||
\end{equation}

There exists an analytical solution (pseudo-inverse matrix) to the optimization problem:

\begin{equation}
\label{equ:pseudo-inv}
\bm{\beta} = (\bm{X^T}\bm{X})^{-1} \bm{X^T} \bm{Y}
\end{equation}

In principle, we can apply Equation (\ref{equ:pseudo-inv}) directly to the non-zero pixels on the binary map to solve for the parameters. However, this method can only utilize the "0" and "1" of each pixel and not exploit the information thoroughly. Besides, it is not helpful to back-propagate through the binary segmentation branch. Therefore, we use the WLS fitting and introduce the weights into Equation (\ref{equ:linear-system}):

\begin{gather}
\bm{WX\beta} = \bm{WY} \\
where \quad \bm{W} = \text{diag}(w_1,w_2,...,w_m) = 
\begin{pmatrix}
w_1 & \cdots & 0 \\
\vdots & \ddots & \vdots \\
0 & \cdots & w_m
\end{pmatrix} \nonumber
\end{gather}

\noindent $w_i$ is the amplitude of the i-th pixel on the weight map before binarization in the binary segmentation branch. As a result, the equation to solve for the fitting parameters becomes:

\begin{equation}
\label{equ:weighted-inv}
\bm{\beta} = (\bm{(WX)^T}\bm{WX})^{-1} \bm{(WX)^T} \bm{WY}
\end{equation}

For each track, we generate (pixel, weight) pairs after the pixel assignment. The WLS of multiple tracks can be computed together in the deep learning framework and it improves the efficiency to a great extent.

To judge the accuracy of WLS fitting, we need to define the loss function. An intuitive method is to use the square errors between the fitting parameters and the label. However, the impacts of the intercept and the slope have different scales, and they work together as a whole. To separate them manually is inappropriate. Hence, we define the geometric loss function as follows:

\begin{equation}
\label{equ:geo-loss}
L = \sum\limits_{x=0}^{W-1} \left[ (\beta_2 x + \beta_1) - (\hat{\beta}_2 x + \hat{\beta}_1) \right]^2
\end{equation}

\noindent where $\beta_1$ and $\beta_2$ are predictions by the WLS fitting, $\hat{\beta}_2$ and $\hat{\beta}_1$ are ground-truth values in the label, and $W$ is the width of the image.

According to Equation (\ref{equ:geo-loss}), it is very convenient to compute the partial derivatives of the loss with regard to the fitting parameters. The core problem is how to propagate the errors to the binary segmentation branch. In Equation (\ref{equ:weighted-inv}), the matrix manipulations are differentiable \cite{RN155}, so we can use the derivation method in the linear algebra to compute the derivatives of $\beta$ with regard to $W$. Through Cholesky decomposition and other available methods, the derivation can be implemented in the deep learning frameworks. So far, we have cleared the last barrier of back-propagation. The constructed end-to-end model can be optimized as a whole.

\subsection{Architectural configurations}

The architecture discussed above is mainly designed for images with multiple and variable tracks. However, it is flexible with different configurations for both the single-track case and the multi-track case.

\begin{table}[htb]
	\caption{\label{tab:conf} Different configurations of the network and their application scenarios.}
	\centering
	%% \tablesize{} %% You can specify the fontsize here, e.g., \tablesize{\footnotesize}. If commented out \small will be used.
	\scriptsize
	\begin{tabular}{ccccc}
		\hline
		\textbf{Configuration}	& \textbf{Network Part}	& \textbf{Output} & \textbf{Loss Function} & \textbf{App. Scenario} \\
		\hline
		Conf (1) & \makecell[c]{Base Network\\Binary Seg.} & \makecell[c]{Hough Out\\Single WLS Out} & Binary Seg. Loss & \makecell[c]{Single Track\\(Coarse)} \\
		\hline
		Conf (2) & \makecell[c]{Base Network\\Binary Seg.} & \makecell[c]{Hough Out\\Single WLS Out} & \makecell[c]{Binary Seg. Loss\\Single WLS Loss} & \makecell[c]{Single Track\\(Precise)} \\
		\hline
		Conf (3) & \makecell[c]{Base Network\\Binary Seg.\\Semantic Seg.} & \makecell[c]{Hough Out\\WLS Out} & \makecell[c]{Binary Seg. Loss\\Semantic Seg. Loss} & \makecell[c]{Multiple Tracks\\(Not End-to-End)} \\
		\hline
		Conf (4) & \makecell[c]{Base Network\\Binary Seg.\\Semantic Seg.} & \makecell[c]{Hough Out\\WLS Out} & \makecell[c]{Binary Seg. Loss\\Semantic Seg. Loss\\WLS Loss} & \makecell[c]{Multiple Tracks\\(End-to-End)} \\
		\hline
	\end{tabular}
\end{table}

Table \ref{tab:conf} summarizes four common configurations and their application scenarios. Conf (1) and Conf (2) are used for the single-track case, while Conf (3) and Conf (4) are used for the multi-track case. Compared to Conf (1), Conf (2) adds the single-track WLS loss and provides the information of the single track with higher precision. Conf (3) and Conf (4) use the whole network including the semantic part. In Conf (3), no WLS loss is used, but the WLS output can still give meaningful results. Compared to Conf (3), the WLS loss is used in Conf (4), which makes it a fully end-to-end model.

\section{Simulation and experimental results}

\subsection{Physical simulation}

To simulate the track projection images in real conditions and provide a basis for comparison between empirical methods and the neural network, we use Garfield++ \cite{RN171}, ROOT \cite{BRUN199781} and other software to produce high energy ion events. The whole process from the injection of ions to the collection of ionization electrons is simulated in the experimental setup discussed in Section \ref{sec:experimental-setup}. Only one projection plane with the pixel sensor is considered for simplicity.

First, configuration files related to the physical process are generated. We use SRIM \cite{ZIEGLER20101818} to compute the statistical characteristics of GeV ions (Kr with 85.9 relative atomic mass in this simulation) passing through the air. The air is represented by gas mixtures of 76\% N$_2$, 23\% O$_2$ and 1\% Ar. The computed characteristics include energy loss rate, projected range and longitudinal/lateral straggling. Besides, we use Magboltz (integrated in Garfield++) to compute the gas table with identical mixtures in the electric field ranging from 0 to 400 V/cm.

\begin{figure}[htb]
	\centering
	\subfigure[]{
		\includegraphics[width=0.45\textwidth]{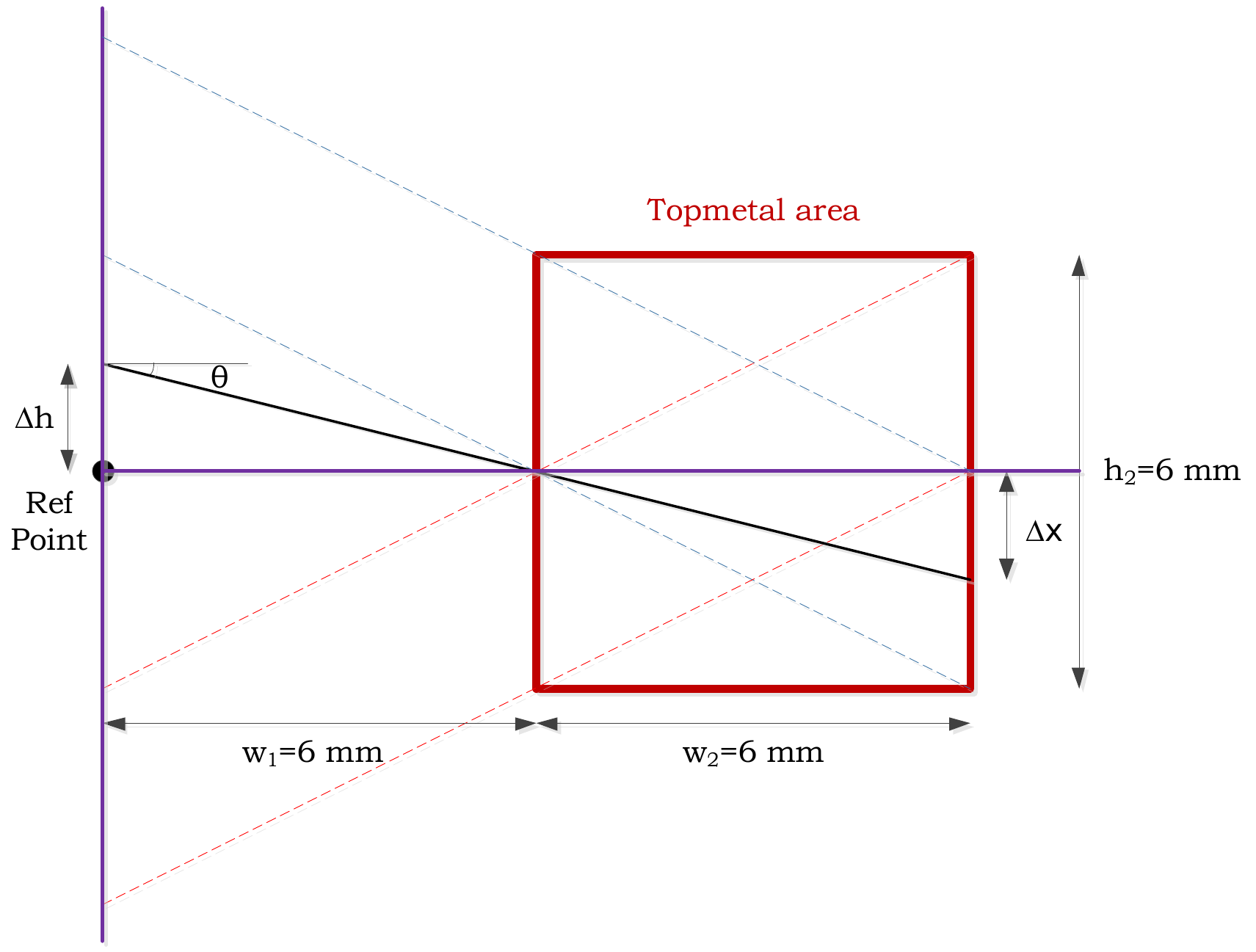}
		%\caption{fig1}
	}
    \subfigure[]{
    	\includegraphics[width=0.45\textwidth]{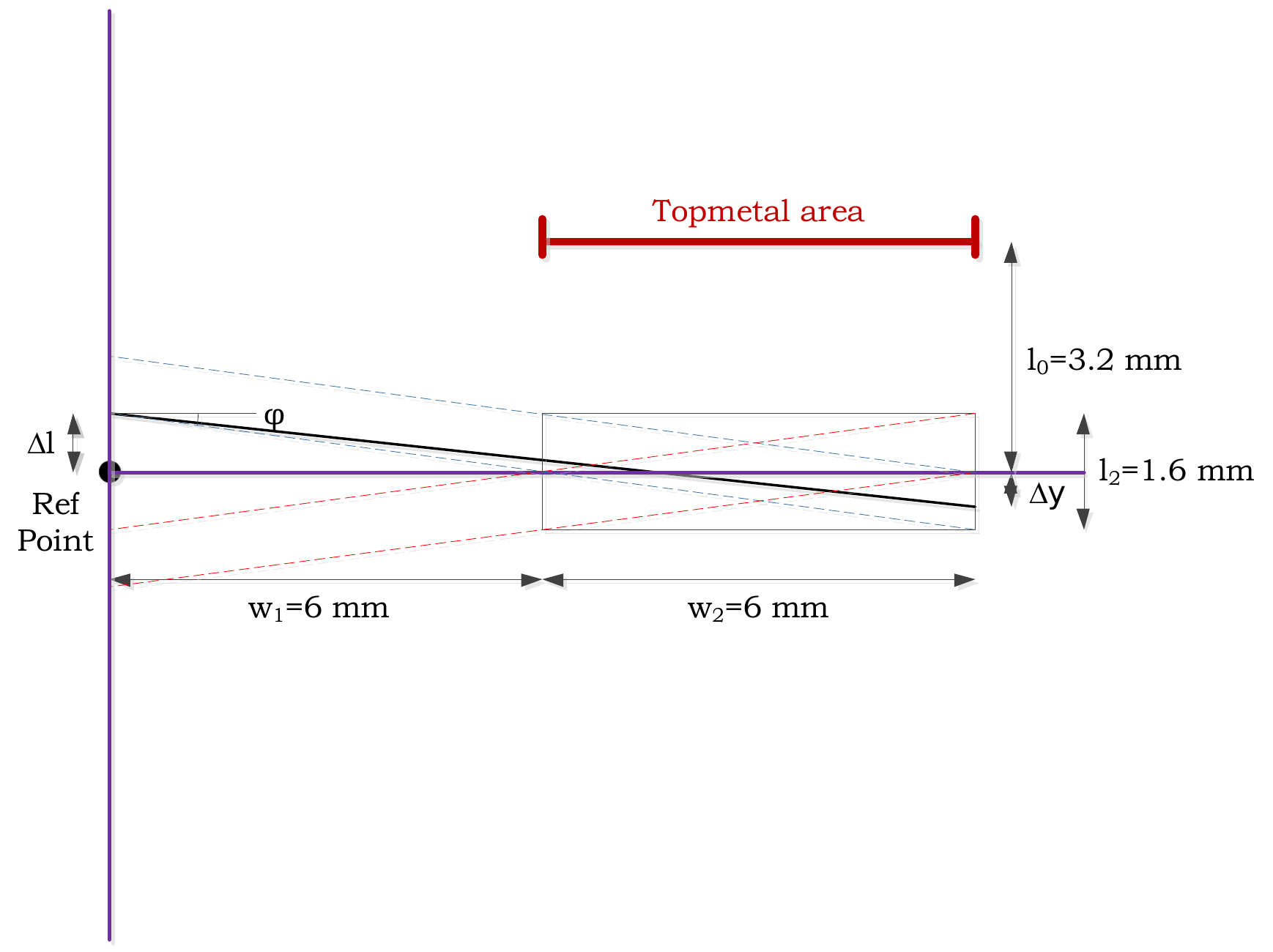}
    	%\caption{fig1}
    }
	\caption{\label{fig:geometry} The geometric structure defined in the Garfield++. The \emph{Topmetal} is placed vertically in the x-z coordinate plane, and ionization electrons drift along the y-axis. (a) A side-view of the detector geometry. (b) A top-view of the detector geometry.}
\end{figure}

Second, the geometric structure, the electric field and the readout sensor are defined. We adopt the geometry of the first detecting cell in Fig. \ref{fig:multi-track-system}, where the ionization electrons drift in the horizontal electric field towards one side, and the \emph{Topmetal} is located vertically. Fig. \ref{fig:geometry} gives the side-view and top-view of the defined geometric structure, and the possible ranges of the tracks. The \emph{Topmetal} plane is set to be the readout sensor.

Third, the paths of ions and ionization clusters are generated. The initial energy of the Kr ion is fixed at 2.0087 GeV. The W value (average energy loss for a single ionization electron) is 30 eV \cite{Valentine_1958}, and the Fano factor is set to 0.3. The ionization electrons appear in clusters with an average of 500 electrons in each cluster. We use a stratified sampling strategy to specify the initial angle and position of ions: the angle is divided into many equal ranges, and equal number of examples is generated in each range considering the randomness; the initial position is constrained by the range and sampled thereafter. This stratified sampling strategy can ensure the coverage of different angles, which is vital to train robust models. After determining the initial kinetics of the ion, we generate tracks and associated clusters of electrons in the framework of Garfield++.

Finally, electrons are drifted to the detecting plane and collected. The collection efficiency is defined as 0.11$\sim$0.14. For electrons in the clusters, we drift them one by one to get their end-state locations. Lateral and longitudinal diffusion is considered in the process. For collected electrons, we fill them into a two-dimensional histogram as the track projection image.

\subsection{Data preprocessing}

In the preprocessing stage, we consider the digital readout of the \emph{Topmetal} sensor, and generate input images and label data/images for the neural network model.

When producing images for the input, we define \emph{foreground} to be tracks in the physical simulation, and \emph{background} to be noise intrinsic in each pixel. The count of electrons in the histogram is converted to sampling values of analog-to-digital converters (ADC) at 16.76 electrons per unit. Multiple histograms are summed in the foreground to generate images with multiple tracks. To avoid the same foreground image being synthesized twice in the training and test dataset, we record the indexes of each generated foregrounds and drop the combination if there is a duplication. For background, according to statistics at extremely low event rates, the noise of each pixel is sampled from a Gaussian distribution with -0.695 mean and 1.701 standard deviation. Finally, the foreground and background are added together. It should be noted that ADC sampling values are usually discrete by nature; however, the values on the synthetic image are continuous. In practice, the projection images from experiments are also continuous because the actual value is the difference between the observed value and the \emph{pedestal} estimated by an average in the idle state without events.

To produce labels for the binary segmentation and semantic segmentation, we take out the intercept and slope along with the histogram. A necessary step is to convert the coordinate system from physical simulation to the convention of images. For binary segmentation, we process the label image column by column, and set the pixel nearest to each straight line to value "1", otherwise value "0". For semantic segmentation, the label image is also processed in the order of columns. First, for 3 points nearest to each straight line in each column, we mark them \emph{On}. The upper 3 points and lower 3 points with regard to \emph{On} are candidates of \emph{Up} and \emph{Down}. If a point is both \emph{Up} and \emph{Down} candidates for different straight lines, we mark it \emph{Both} in priority. The remaining candidates are marked as \emph{Up} or \emph{Down}. Finally, other points in the semantic label image are marked as \emph{Off}. If the points are close to the edge of the image, out-of-range points are not marked.

When the above preprocessing work is done, we save the input image, the binary label image and the semantic label image to files along with the intercept and slope of each track.

\subsection{Configurations}

For single-track and multi-track simulations, we use 10,000 examples for the training dataset and 2,000 examples for the test dataset. The L2 loss is used for regularization. We choose the stochastic gradient descent with momentum as the optimization algorithm. The initial learning rate and the momentum coefficient are set to 0.0001 (except for 0.000005 in the multi-track end-to-end finetuning) and 0.9. The batch size is 4 in single-track or not end-to-end conditions. When training the end-to-end neural network for the multi-track with the WLS loss, the batch size is 1. At different stages, we train from randomly initialized weights or on the basis of former weights (discussed in following sections). The variance scaling initializer \cite{7410480} is used for weights in the convolution layers. For weights in deconvolution layers, we compute a standard deviation according to the number of input channels, and truncated normal initializer (dropping samples out of two sigma) is used with 0 mean and this standard deviation. The biases in convolution layers and deconvolution layers are initialized to 0. We stop training and test our model on the test dataset when the total loss has decreased substantially. The network model is implemented with the TensorFlow \cite{DBLP:conf/osdi/AbadiBCCDDDGIIK16} software on a desktop computer with NVIDIA GeForce TITAN X GPU (12 GB).

\subsection{Single-track simulation}
\label{sec:single-sim}

\begin{figure}[htb]
	\centering
	\includegraphics[width=0.85\textwidth]{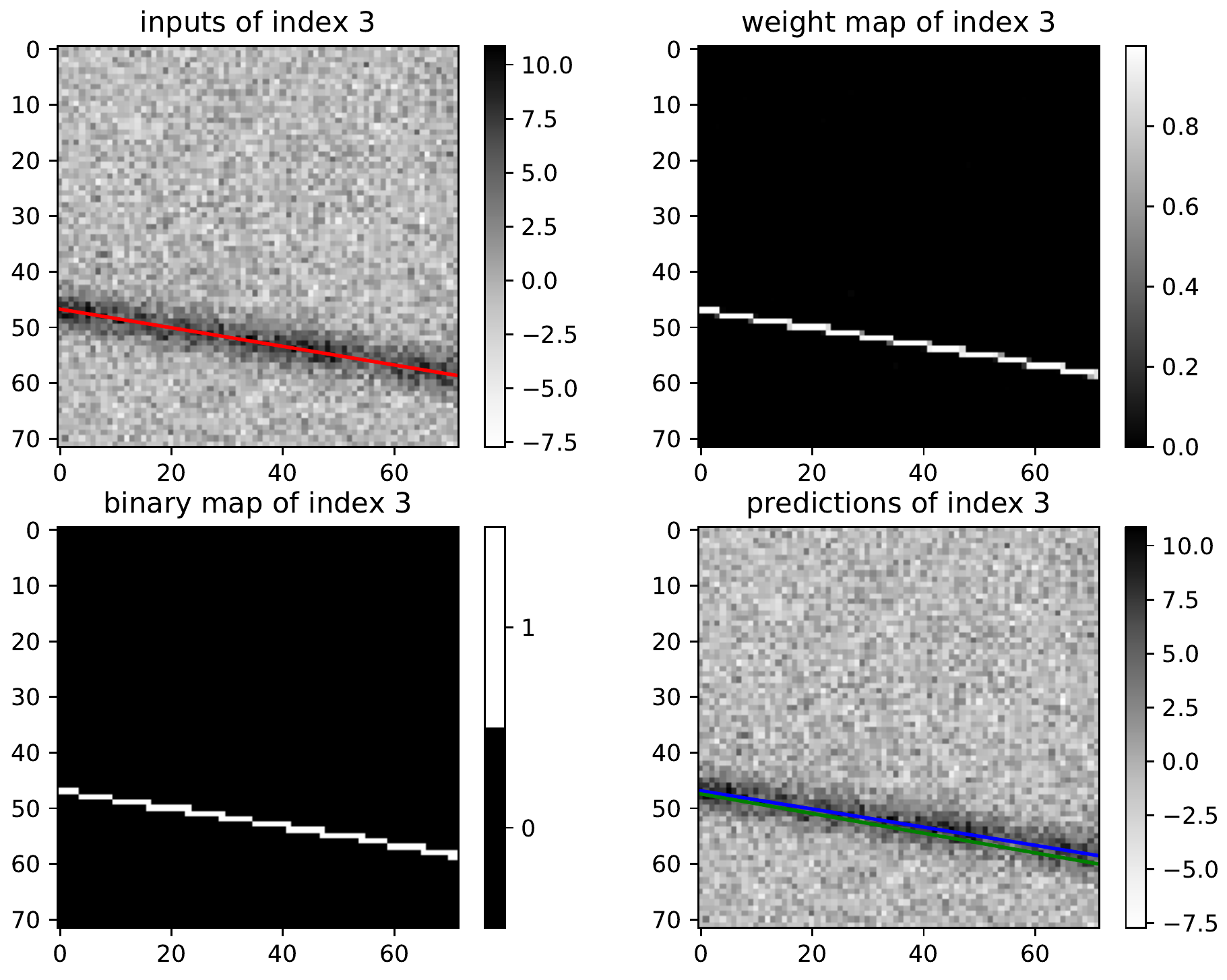}
	\caption{\label{fig:single-signoff-ind3} An example of images at different stages of the neural network model for the single-track simulation. (left top) The original input with the label (red line). (right top) The weight map before binarization in the binary segmentation branch. (left bottom) The binary map. (right bottom) Predictions of the neural network model (blue line: WLS fitting; green line: Hough output).}
\end{figure}

\begin{table}[htb]
	\caption{\label{tab:single-int-slp} The statistics of residuals of intercept and slope predictions. For the single-track image, five methods are analyzed. The spatial resolution is estimated based on the pixel size of the \emph{Topmetal-II-} chip in \cite{AN2016144}.}
	\centering
	%% \tablesize{} %% You can specify the fontsize here, e.g., \tablesize{\footnotesize}. If commented out \small will be used.
	\scriptsize
	\begin{tabular}{ccccccc}
		\hline
		\multirow{2}{*}{\textbf{Method}} & \multicolumn{3}{c}{\textbf{Residual of Intercept}} & \multicolumn{3}{c}{\textbf{Residual of Slope}} \\
		\cline{2-7}
		& mean & std. & resolution (\si{\micro\meter}) & mean & std. & resolution (degree) \\
		\hline
		double edge detection & 0.17711 & 0.62096 & 51.664 & 0.00009 & 0.00828 & 0.474 \\
		mass center & -0.00656 & 0.17119 & 14.243 & 0.00017 & 0.00393 & 0.225 \\
		Hough output & 0.58766 & 0.41500 & 34.528 & 0.00018 & 0.01002 & 0.574 \\
		WLS (w/o loss) & -0.01342 & 0.11147 & 9.274 & -0.00002 & 0.00278 & 0.159 \\
		WLS (w/ loss) & -0.00087 & 0.10559 & 8.785 & 0.00000 & 0.00260 & 0.149 \\
		\hline
	\end{tabular}
\end{table}

\begin{table}[htb]
	\caption{\label{tab:single-cam} Statistics of center measure, angle measure and CAM. For the single-track image, six methods are analyzed.}
	\centering
	%% \tablesize{} %% You can specify the fontsize here, e.g., \tablesize{\footnotesize}. If commented out \small will be used.
	\scriptsize
	\begin{tabular}{cccc}
		\hline
		\textbf{Method}	& \textbf{Center Measure} & \textbf{Angle Measure} & \textbf{CAM} \\
		\hline
		double edge detection & 0.9849 $\pm$ 0.0069 & 0.9961 $\pm$ 0.0030 & 0.9811 $\pm$ 0.0077 \\
		mass center & 0.9984 $\pm$ 0.0034 & 0.9985 $\pm$ 0.0018 & 0.9970 $\pm$ 0.0045 \\
		Hough output (original) & 0.9938 $\pm$ 0.0072 & 0.9951 $\pm$ 0.0033 & 0.9790 $\pm$ 0.0080 \\
		Hough output (adjusted) & 0.9941 $\pm$ 0.0042 & 0.9951 $\pm$ 0.0033 & 0.9892 $\pm$ 0.0052 \\
		WLS (w/o loss) & 0.9988 $\pm$ 0.0010 & 0.9987 $\pm$ 0.0011 & 0.9975 $\pm$ 0.0016 \\
		WLS (w/ loss) & 0.9989 $\pm$ 0.0009 & 0.9988 $\pm$ 0.0010 & 0.9977 $\pm$ 0.0014 \\
		\hline
	\end{tabular}
\end{table}

For the single-track simulation dataset, we create the network architecture described in Section \ref{sec:arch} for training and testing, and compare the results with the traditional empirical methods in Section \ref{sec:empirical}. We use the base network and the binary segmentation part (Conf (1) and Conf (2) in Table \ref{tab:conf}). First, we do not back-propagate through the WLS loss and perform relevant training and testing. Then the WLS loss is used to finetune the whole network.

In Fig. \ref{fig:single-signoff-ind3}, we give an example showing how the network model deals with the single-track image. It can be seen in the weight map that the amplitudes of pixels are highly centralized; most pixels have values close to zero, and only pixels on the center line of the track have values close to one. In the binary map, the features become even clearer. By carefully examining the predictions of the neural network model, we can see a small shift (mainly the intercept) of the Hough transform output compared to the label. This non-ideal phenomenon is due to the discrete Hough space. In Table \ref{tab:single-cam}, we add an adjustment to the Hough output to correct the shift and give results before and after the adjustment.

To quantify the performance of different methods, based on the intercept and slope predictions on the test dataset, we fit a Gaussian distribution to the residuals (difference from label) and list the mean and standard deviation (std.) in Table \ref{tab:single-int-slp}. The mean values represent the system bias of different methods, which could be greatly reduced by adding/subtracting a constant. The standard deviation could represent the resolution achieved by the method, which is also listed beside the standard deviation. We exclude some outliers when doing statistics of the double edge detection method. When using WLS (w/ loss), the resolution for the initial position (intercept) and angle (slope) could be 8.785 \si{\micro\meter} and 0.149$^{\circ}$, which is the best achievable result.

In the above analysis, we deal with the initial position and angle separately. However, they are two sides of a single problem and cannot be totally isolated. To quantify the precision in a systematic way, we introduce the \emph{center-angle measure} (CAM) which combines both:

\begin{gather}
c = 1 - \frac{||\bm{C}_{label} - \bm{C}_{pred}||_2}{0.5 \cdot L_{label}} \\
a = 1 - \frac{|\theta_{label} - \theta_{pred}|}{90^{\circ}} \\
CAM = c \cdot a
\end{gather}

\noindent where $\bm{C}_{label}$ and $\bm{C}_{pred}$ represent the center positions of the label line segment and the predicted line segment. We calculate the L2 distance between the two centers and divide it by half the length of the label before subtraction. $\theta_{label}$ and $\theta_{pred}$ represent the angles of the label line and the predicted line. We calculate the absolute difference between the two angles and divide it by 90 degrees before subtraction. The center positions and angles are converted from intercepts and slopes. The product of the center measure and angle measure is defined as the CAM.

In Table \ref{tab:single-cam}, we gather the results of center measure, angle measure and CAM using different methods on the test dataset. It is consistent with the results in Table \ref{tab:single-int-slp}. The CAM achieved by WLS (w/ loss) is 0.9977 $\pm$ 0.0014 and indicates superior precision. In comparison, the CAM of the mass center is 0.9970 $\pm$ 0.0045, which has smaller mean value and much larger fluctuations.

\begin{figure}[htb]
	\centering
	\includegraphics[width=0.48\textwidth]{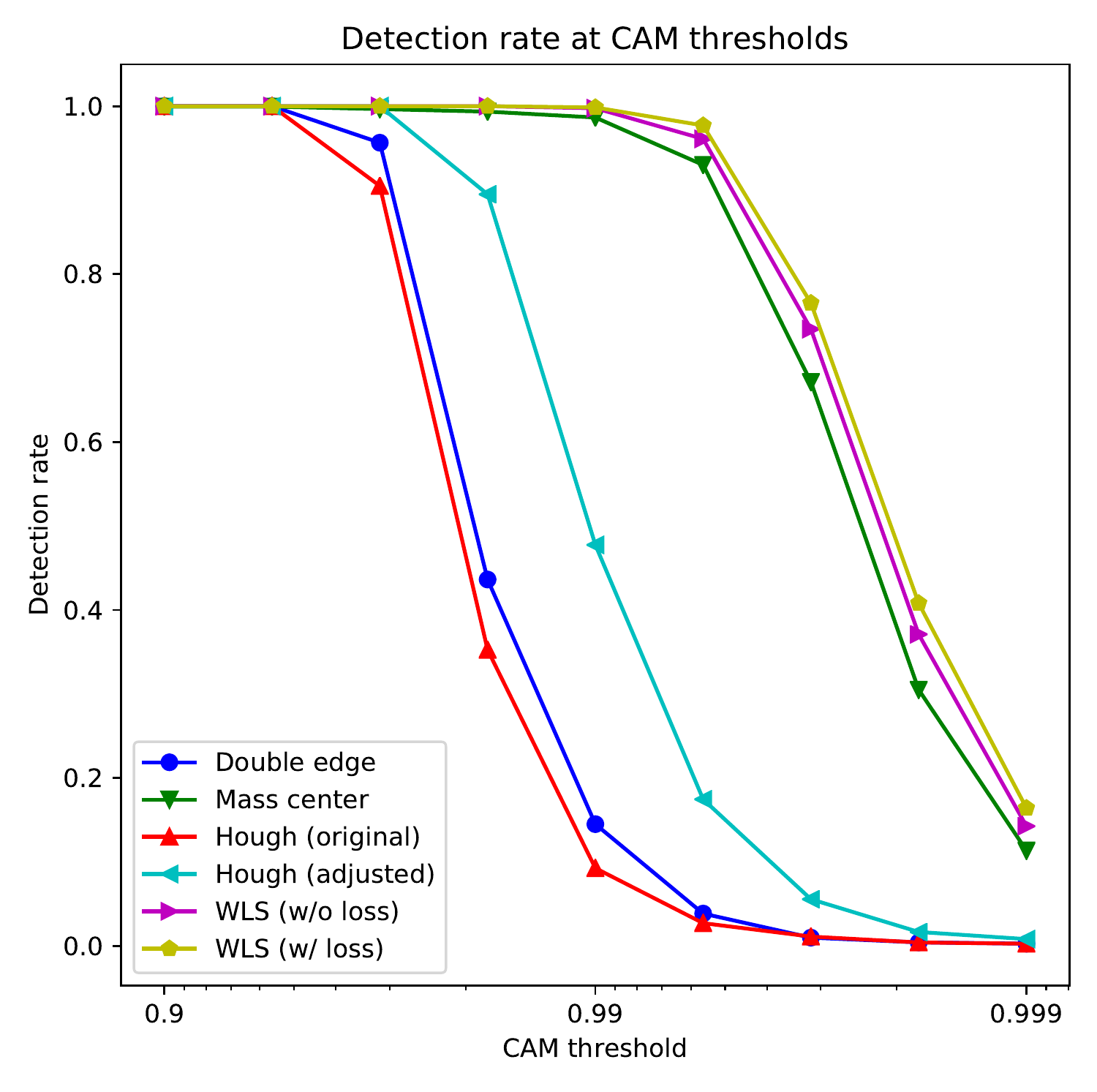}
	\caption{\label{fig:single-track-det-rate} Detection rates of different methods versus the CAM threshold for the single-track image. At a specific threshold, the test results with CAM above the threshold are considered as detected. We use the logarithmic x-axis (from 0.9 to 0.999) in the figure.}
\end{figure}

In Fig. \ref{fig:single-track-det-rate}, we compare the detection rates of different methods at CAM thresholds. The detection rate refers to the ratio of tested images above a CAM threshold. It can be seen that the overall trend of the curves is in good accordance with Table \ref{tab:single-cam}. The difference between the mass center and WLS is not large. However, the advantage of WLS is still noticeable, and the relative improvement is more evident at high CAM thresholds.

\subsection{Multi-track simulation}

\begin{figure}[htb]
	\centering
	\includegraphics[width=0.85\textwidth]{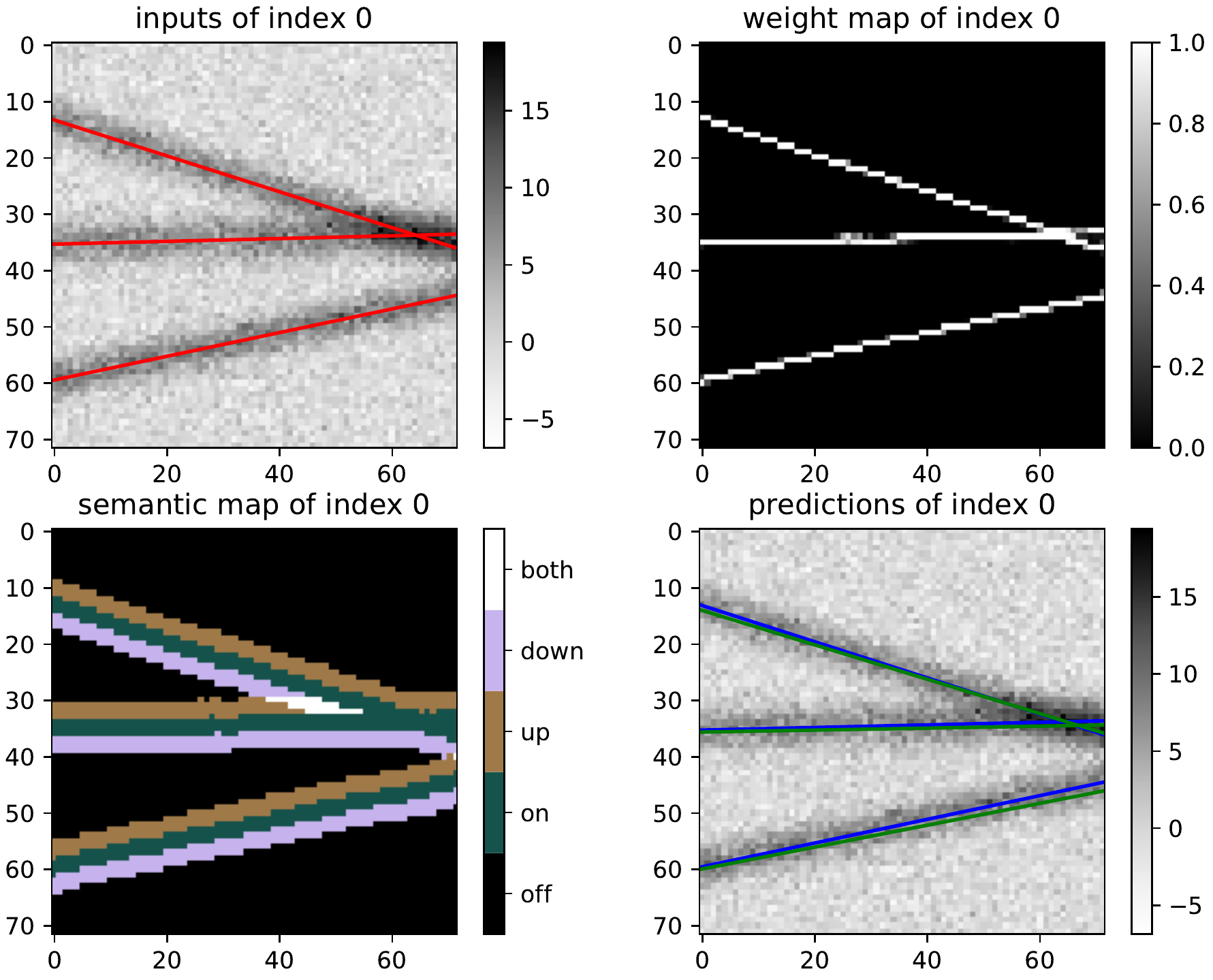}
	\caption{\label{fig:multi-signoff-ind0} An example of images at different stages of the neural network model for the multi-track simulation. (left top) The original input with the label (red line). (right top) The weight map before binarization in the binary segmentation branch. (left bottom) The result of semantic segmentation. (right bottom) Predictions of the neural network model (blue line: WLS fitting; green line: Hough output).}
\end{figure}

\begin{table}[htb]
	\caption{\label{tab:multi-int-slp} The statistics of residuals of intercept and slope predictions. For the multi-track image, three methods are analyzed. The upper part uses 1-3 tracks, and the lower part uses 1-5 tracks. The spatial resolution is estimated based on the pixel size of the \emph{Topmetal-II-} chip in \cite{AN2016144}.}
	\centering
	%% \tablesize{} %% You can specify the fontsize here, e.g., \tablesize{\footnotesize}. If commented out \small will be used.
	\scriptsize
	\begin{tabular}{ccccccc}
		\hline
		\multirow{2}{*}{\textbf{Method}} & \multicolumn{3}{c}{\textbf{Residual of Intercept}} & \multicolumn{3}{c}{\textbf{Residual of Slope}} \\
		\cline{2-7}
		& mean & std. & resolution (\si{\micro\meter}) & mean & std. & resolution (degree) \\
		\hline
		Hough output & 0.61478 & 0.43175 & 35.922 & 0.00001 & 0.01036 & 0.594 \\
		WLS (w/o loss) & 0.01227 & 0.15250 & 12.688 & 0.00022 & 0.00403 & 0.231 \\
		WLS (w/ loss) & 0.00419 & 0.13722 & 11.417 & 0.00003 & 0.00366 & 0.210 \\
		\hline
		Hough output & 0.60142 & 0.44297 & 36.855 & -0.00014 & 0.01086 & 0.622 \\
		WLS (w/o loss) & 0.00916 & 0.19269 & 16.032 & 0.00047 & 0.00553 & 0.317 \\
		WLS (w/ loss) & -0.01206 & 0.18281 & 15.210 & 0.00011 & 0.00504 & 0.289 \\
		\hline
	\end{tabular}
\end{table}

\begin{table}[htb]
	\caption{\label{tab:multi-cam} Statistics of center measure, angle measure and CAM. For the multi-track image, four methods are analyzed. The upper part uses 1-3 tracks, and the lower part uses 1-5 tracks.}
	\centering
	%% \tablesize{} %% You can specify the fontsize here, e.g., \tablesize{\footnotesize}. If commented out \small will be used.
	\scriptsize
	\begin{tabular}{cccc}
		\hline
		\textbf{Method}	& \textbf{Center Measure} & \textbf{Angle Measure} & \textbf{CAM} \\
		\hline
		Hough output (original) & 0.9833 $\pm$ 0.0075 & 0.9949 $\pm$ 0.0035 & 0.9783 $\pm$ 0.0081 \\
		Hough output (adjusted) & 0.9939 $\pm$ 0.0044 & 0.9949 $\pm$ 0.0035 & 0.9888 $\pm$ 0.0056 \\
		WLS (w/o loss) & 0.9983 $\pm$ 0.0018 & 0.9983 $\pm$ 0.0017 & 0.9966 $\pm$ 0.0030 \\
		WLS (w/ loss) & 0.9985 $\pm$ 0.0016 & 0.9984 $\pm$ 0.0016 & 0.9968 $\pm$ 0.0026 \\
		\hline
		Hough output (original) & 0.9836 $\pm$ 0.0077 & 0.9947 $\pm$ 0.0040 & 0.9784 $\pm$ 0.0087 \\
		Hough output (adjusted) & 0.9936 $\pm$ 0.0048 & 0.9948 $\pm$ 0.0038 & 0.9884 $\pm$ 0.0064 \\
		WLS (w/o loss) & 0.9977 $\pm$ 0.0033 & 0.9977 $\pm$ 0.0028 & 0.9954 $\pm$ 0.0054 \\
		WLS (w/ loss) & 0.9979 $\pm$ 0.0030 & 0.9979 $\pm$ 0.0026 & 0.9958 $\pm$ 0.0050 \\
		\hline
	\end{tabular}
\end{table}

For the multi-track simulation dataset, we create the full network architecture comprised of the base network, binary segmentation and semantic segmentation (Conf (3) and Conf (4) in Table \ref{tab:conf}). First, training and testing are performed without the WLS loss. Next, the loss is used to finetune the network end-to-end.

In Fig. \ref{fig:multi-signoff-ind0}, we give an example visualizing different stages of the network. The left bottom image is the result from semantic segmentation, and other images and their annotations are the same as Fig. \ref{fig:single-signoff-ind3}. The multi-track discrimination is more difficult than the single-track case. In this example, two tracks are close to each other, and intersect at the right edge of the input image. In the weight map, the intersecting region is a little vague; however, the "X" shape is well distinguishable. In the semantic image, there are also some non-ideal pixels, but the overall condition is fairly good. The network predictions are approaching the label. Again, the Hough output has a shift compared to the WLS output, which could be fixed with an adjustment.

To set up relations between the predicted tracks and tracks in the label, we use a \emph{greedy match algorithm}. For each example in the test dataset, a CAM matrix is built between predicted tracks and label tracks, with each element being the CAM calculated using the prediction and the label. Each time we select the maximum element in the CAM matrix, record the relation and leave out the row and the column in subsequent matching. The process continues until no elements in the CAM matrix satisfying the condition.

In Table \ref{tab:multi-int-slp} and Table \ref{tab:multi-cam}, we show the statistical results of Hough output and WLS in different conditions. Two multi-track datasets are used in this section: 1-3 tracks dataset and 1-5 tracks dataset. They are listed in the upper part and low part of the tables, respectively. Again WLS (w/ loss) has the best precision. It should be noted that the resolution achieved by WLS (w/ loss) in 1-3 tracks dataset (11.417 \si{\micro\meter} and 0.210$^{\circ}$) is still better than the mass center method in the single-track dataset (14.243 \si{\micro\meter} and 0.225$^{\circ}$). Besides, when using the 1-5 tracks dataset, the performance only deteriorates a little. For CAM results, the trend is almost the same.

\begin{figure}[htb]
	\centering
	\subfigure[]{\label{fig:ext-f1-macro} \includegraphics[width=0.48\linewidth]{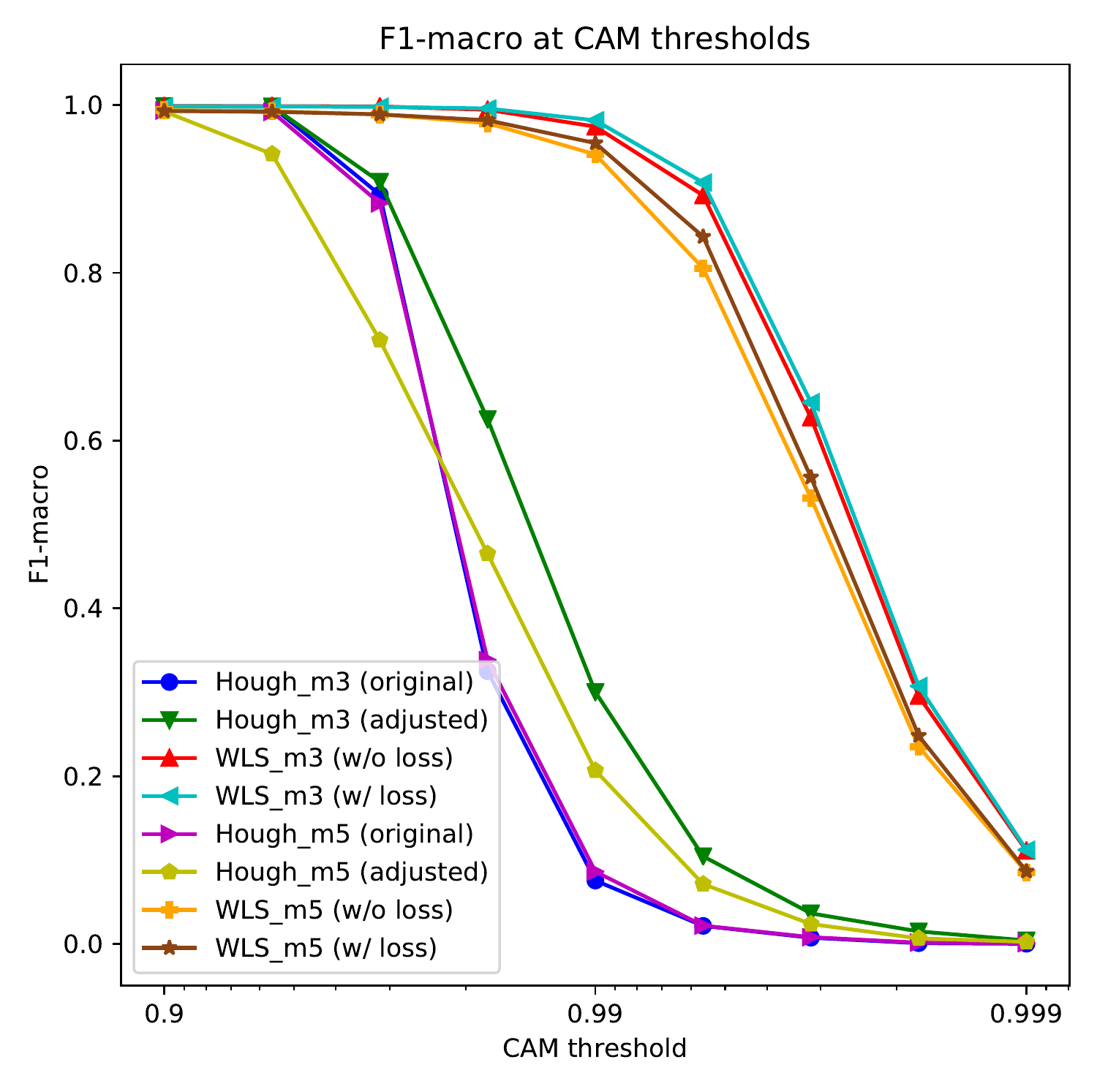}}
	\subfigure[]{\label{fig:ext-f1-micro} \includegraphics[width=0.48\linewidth]{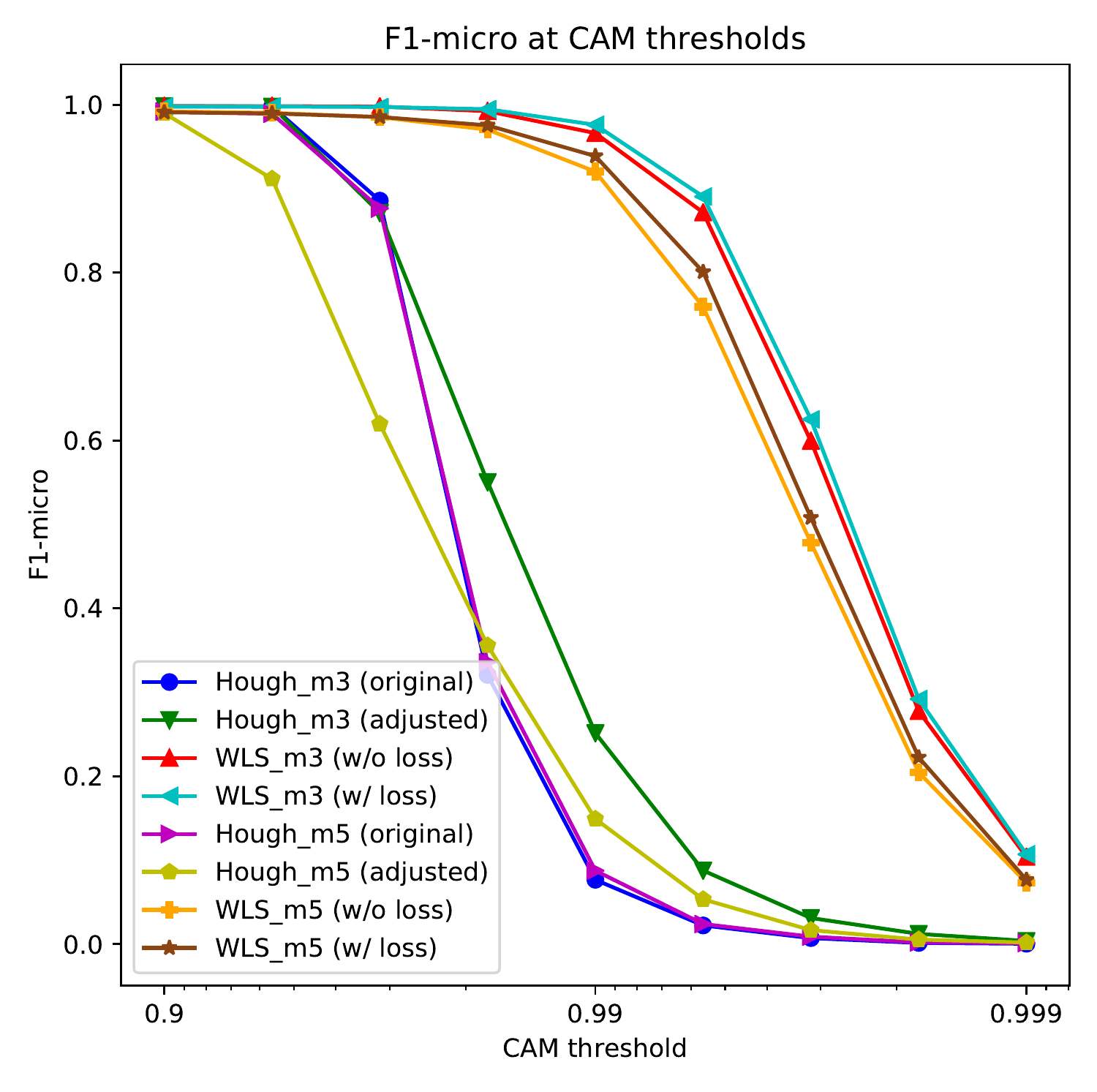}}
	\caption{The suffix "m3" means dataset with 1-3 tracks, and the suffix "m5" means dataset with 1-5 tracks. We use the logarithmic x-axis (from 0.9 to 0.999) in these figures. (a) F1-macro of different methods versus the CAM threshold for the multi-track image. F1-macro is the average of F1 scores of individual examples. (b) F1-micro of different methods versus the CAM threshold for the multi-track image. F1-micro is the F1 score of all the examples.}
\end{figure}

Finally, we give the F1-macro and F1-micro curves of different methods versus the CAM threshold. We define $TP_i$ to be successfully matched tracks above the CAM threshold in an example, $(TP_i + FP_i)$ to be total predicted tracks in an example and $(TP_i + FN_i)$ to be total label tracks in an example. The F1-macro and F1-micro can thus be calculated:

\begin{gather}
P_i = \frac{TP_i}{TP_i + FP_i}, \quad
R_i = \frac{TP_i}{TP_i + FN_i}, \quad
F1_i = \frac{2 \cdot P_i \cdot R_i}{P_i + R_i} \\
\text{F1-macro} = \frac{1}{N} \sum\limits_{i=1}^N F1_i
\end{gather}

\begin{gather}
P = \frac{\sum\limits_{i=1}^N TP_i}{\sum\limits_{i=1}^N (TP_i + FP_i)}, \quad
R = \frac{\sum\limits_{i=1}^N TP_i}{\sum\limits_{i=1}^N (TP_i + FN_i)} \\
\text{F1-micro} = \frac{2 \cdot P \cdot R}{P + R}
\end{gather}

The curves of F1-macro and F1-micro are shown in Fig. \ref{fig:ext-f1-macro} and Fig. \ref{fig:ext-f1-micro}. It can be seen that the WLS fitting is better than Hough output in both figures, which is in good accordance with Table \ref{tab:multi-cam}. The WLS (w/ loss) is slightly better than the WLS (w/o loss), and it is more obvious in the F1-micro figure. The results with 1-5 tracks are slightly worse than those with 1-3 tracks because of the track images are more complicated. For the Hough output, we notice a drop of the Hough output (adjusted) for the 1-5 tracks dataset at low CAM thresholds. Preliminarily we attribute the drop to the greedy match algorithm we use.

\subsection{Analysis of experimental images}
\label{sec:exp-images}

\begin{figure}[htb]
	\centering
	\includegraphics[width=0.95\textwidth]{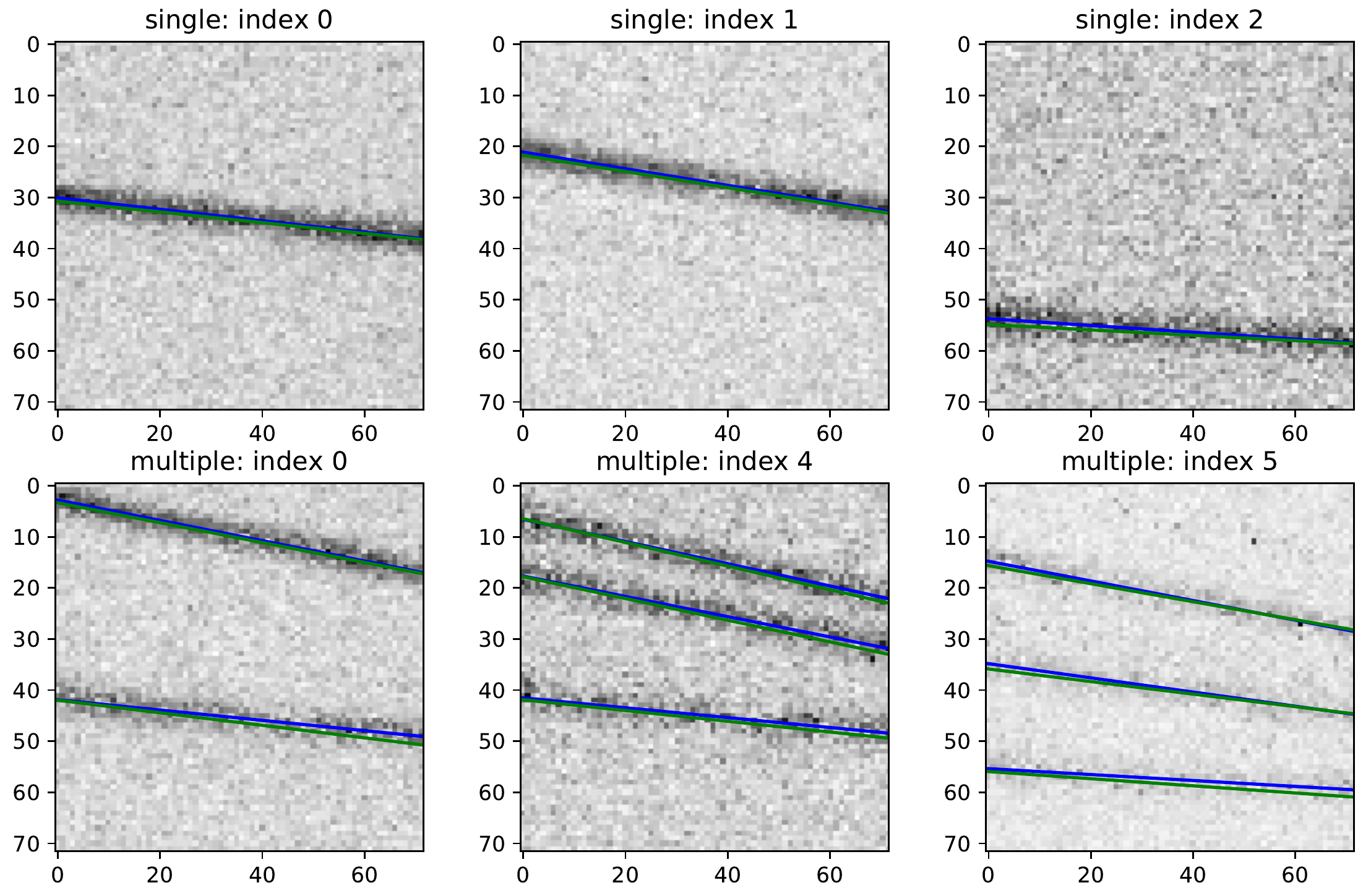}
	\caption{\label{fig:exp-signoff} Test results of applying the trained network weights to experimental data. The upper three images show the single-track case, and the lower three images show the multi-track case. The predictions of WLS fitting (blue line) and Hough output (green line) are plotted.}
\end{figure}

To demonstrate the effectiveness of the proposed network architecture in experiments, we directly apply the trained network weights to the experimental images and conduct a qualitative study. These images come from the experiment in \cite{LI2020163697}. The results are shown in Fig. \ref{fig:exp-signoff}. In these figures, the gray-scale background image is the experimental data, and the line segments are predictions of the WLS fitting and the Hough output. Based on the observation, the divergence of the background track images is considerably large. For example, in the right top image, the difference between the noisy background and the track is relatively small, which indicates the amplitude change on the track is not obvious; in the right bottom image, the pattern of tracks is very centralized, which may be caused by the short drifting distance of the ionization electrons. In spite of these diversities and discrepancy between simulations and experiments, the weights trained with the simulation data could very well recognize the tracks in the experiment. Whether the WLS fitting or the Hough output is used, it can correctly judge the number of tracks and give the accurate position and angle. When we carefully examine the images, we find that the prediction of WLS fitting is nearer to the center line of each track. This is consistent with the previous simulations.

\section{Conclusions}

In this paper, we mainly discuss the multi-track location and orientation problem in the gaseous drift chamber based on the \emph{Topmetal} series pixel sensor. First, we briefly introduce the detecting instrument used in the previous experiment and the conceptual design implemented with the developing \emph{Topmetal-CEE} chip. When describing the scheme of measurement, we emphasize on the capability of multi-track detection enabled by the time \& amplitude readout of the \emph{Topmetal-CEE}. Next, two traditional empirical methods are introduced. Their limitations shed light on the direction to design novel deep learning methods. The proposed architecture is an end-to-end neural network based on segmentation and fitting. The network architecture could be trained end-to-end to eliminate irretrievable errors from stage to stage. In the single-track simulation, the WLS fitting outperforms the traditional mass center method, and the result can be further improved by back-propagating through the WLS loss. In the multi-track simulation, the proposed network model works steadily, and WLS (w/ loss) shows superior performance. Finally, results on the experimental data demonstrate the robustness of the physical simulation and the validity of the method in real-world conditions.

In the future, we would like to deploy the algorithm on application-specific integrated circuits or field programmable gate arrays for online and on-site feature extraction. To achieve this, the optimization and quantization of the network architecture is necessary.

\section*{Acknowledgements}

This research is partly supported by the National Natural Science Foundation of China (Grant Number 11875146, U1932143), and partly supported by the National Key Research and Development Program of China (Grant Number 2016YFE0100900).

%\bibliography{mybibfile.bib}

\end{document}